\documentclass[12pt,letterpaper]{article}
\pdfoutput=1

\usepackage[
backend=biber,
sorting=none,
style=numeric-comp
]{biblatex}
\usepackage{geometry}
\usepackage{tikz}

\usepackage{color}
\usepackage{slashed}
\definecolor{refkey}{rgb}{1,0,0}
\definecolor{labelkey}{rgb}{0,0,1}
\usepackage{bbold,amsfonts,amssymb,amsmath}
\usepackage{esint}
\usepackage{rotating}
\usepackage[utf8]{inputenc}

\usepackage{graphicx}
\usepackage{caption}
\usepackage{subcaption}
\usepackage{booktabs}

\usepackage{listings}

\usepackage{hyperref}

\newcommand{\be}{\begin{equation}}
\newcommand{\ee}{\end{equation}}
\newcommand{\ben}{\begin{displaymath}}
\newcommand{\een}{\end{displaymath}}
\newcommand{\bea}{\begin{equation}\begin{aligned}}
\newcommand{\eea}{\end{aligned}\end{equation}}
\newcommand{\bean}{\begin{eqnarray*}}
\newcommand{\eean}{\end{eqnarray*}}

\newcommand{\bra}[1]{\mbox{$\langle #1 |$}}
\newcommand{\ket}[1]{\mbox{$| #1 \rangle$}}

\newcommand{\eg}{{\it e.g.}}

\newcommand{\commentout}[1]{}

\newcommand{\beq}{\begin{equation}}
\newcommand{\eeq}{\end{equation}}
\newcommand{\beqr}{\begin{displaymath}}
\newcommand{\eeqr}{\end{displaymath}}
\newcommand{\beqa}{\begin{eqnarray}}
\newcommand{\eeqa}{\end{eqnarray}}
\newcommand{\beqar}{\begin{eqnarray*}}
\newcommand{\eeqar}{\end{eqnarray*}}

\newcommand{\cO}{{\cal O}}
\newcommand{\cF}{{\cal F}}
\newcommand{\cL}{{\cal L}}

\newcommand{\non}{\nonumber}

\newcommand{\half}{\ensuremath{\frac{1}{2}}}
\newcommand{\mpi}{\ensuremath{m_{\pi}}}
\newcommand{\fpi}{\ensuremath{f_{\pi}}}
\newcommand{\bmpi}{\ensuremath{\bar{m}_{\pi}}}

\newcommand{\MeV}{\ensuremath{\mbox{MeV}}}
\newcommand{\GeV}{\ensuremath{\mbox{GeV}}}

\begin{document}

\title{\Large \bf Quark Mass Dependence in Gauge Theory Bootstrap\\}

\author{
	Nick Evans$^\text{1}$, Wanxiang Fan$^\text{1}$, Yifei He$^\text{2}$,
	Martin Kruczenski$^\text{3}$ \thanks{\texttt{N.J.Evans@soton.ac.uk, W.Fan@soton.ac.uk, yifei.he@ens.fr, markru@purdue.edu.}} \\
[2.0mm]
$^1$ \small School of Physics \& Astronomy and STAG Research Centre, \\ \small University of Southampton, Highfield, Southampton S017 1BJ, UK\\
$^2$ \small Laboratoire de Physique de l'\'Ecole Normale Supérieure, ENS, Université PSL,\\
\small CNRS, Sorbonne Université, Université Paris Cité, F-75005 Paris, France \\
$^3$ \small Department of Physics and Astronomy, \\
\small Purdue University, West Lafayette, IN 47907, USA.}

\date{}

\maketitle

\begin{abstract}
 We study the quark mass dependence of the Gauge Theory Bootstrap analysis of QCD. The quark mass enters as a free UV parameter through the scalar sum rules, and is related to the pion mass by the Gell-Mann-Oakes-Renner relation. For the quark mass ranging from $1/8$ to 9 times its physical value, we extract the $\sigma,\rho$ and $f_2$ resonances, compute the low energy effective field theory parameters and analyze their quark mass/pion mass dependence.
 The results compare reasonably well with known lattice and chiral perturbation theory results. In particular, as the pion mass is increased, the $\sigma$-meson pole reaches the real axis on the second sheet, below threshold, and separates into two virtual states.  The $\rho,f_2$-resonance poles reaches the real axis at threshold at which point the mesons would be stable.         


\end{abstract}

\clearpage

\tableofcontents

\newpage

\section{Introduction}

The modern S-matrix bootstrap program \cite{Paulos:2016fap,Paulos:2016but,Paulos:2017fhb,Kruczenski:2022lot} allows numerically mapping out of the space of two-to-two scattering amplitudes, leading to bounds on the low energy parameters such as couplings of generic theories. Although these ideas can be used to study pion physics \cite{Guerrieri:2018uew,Bose:2020cod,Bose:2020shm,Guerrieri:2020bto,Albert:2022oes,Fernandez_2023,Albert:2023jtd,Albert:2023seb,Ma:2023vgc}, the actual pion coupling may not saturate such generic bounds and more information is needed, for example from experiment \cite{Guerrieri:2024jkn}. If one, instead of using phenomenological data, wants to identify the correct theory of pions by the fact that it is the low energy theory of QCD, dynamical information from the UV theory has to be incorporated in the bootstrap. The recently proposed Gauge Theory Bootstrap (GTB) \cite{PhysRevLett.133.191601,PhysRevD.110.096001,He:2024nwd,He:2025gws} realized this by combining the form factor bootstrap \cite{Karateev:2019ymz,Chen:2021pgx,Correia:2022dyp,Cordova:2023wjp} that includes spectral densities and form factors with perturbative QCD information on the local operators, in particular through the SVZ sum rules \cite{SHIFMAN1979385,SHIFMAN1979448,Shifman:1978bw} and asymptotic form factors \cite{osti_1447331,Lepage:1980fj}. These provide powerful constraints on the possible low energy theories that match QCD at high energy. The input parameters of the method are the defining QCD parameters $N_c, N_f, m_q, \alpha_s$ and two low energy parameters $m_\pi, f_\pi$ in the nonlinear sigma model description of pions,
and the output are the scattering phase shifts, spectral densities and form factors. From there, one can extract the resonance states and low energy parameters, and the results agree reasonable well with experiment in the case of $N_f=2, N_c=3$ for physical pion and quark masses \cite{PhysRevLett.133.191601,PhysRevD.110.096001,He:2024nwd,He:2025gws}. To understand how powerful this  procedure is, since it uses some assumptions such as that intermediate energy physics is approximately dominated by two pion states \cite{He:2025gws},  it is useful to explore the space of the UV theory parameters and see if the results are still meaningful. In this paper we chose to change the quark mass $m_q$ (equivalently the pion mass $m_\pi$). This is for two main reasons, one is that many lattice computations are done with heavier pions \cite{Davier:2023cyp} and therefore our results can be potentially compared with simulations. The other is in the other direction, decreasing the pion mass puts us closer to the chiral limit where chiral symmetry is restored. Our numerical methods allow us to explore the region $\frac{1}{2\sqrt{2}}m_\pi^{\text{phys}}\lesssim m_\pi \lesssim 3m_\pi^{\text{phys}}$ where $m^{\text{phys}}_\pi$ denotes the physical pion mass $m^{\text{phys}}_\pi\simeq 140\MeV$ and $m_\pi$ is the varying pion mass in our calculations.  

\begin{figure}[t]
	\centering
	\includegraphics[width=\textwidth]{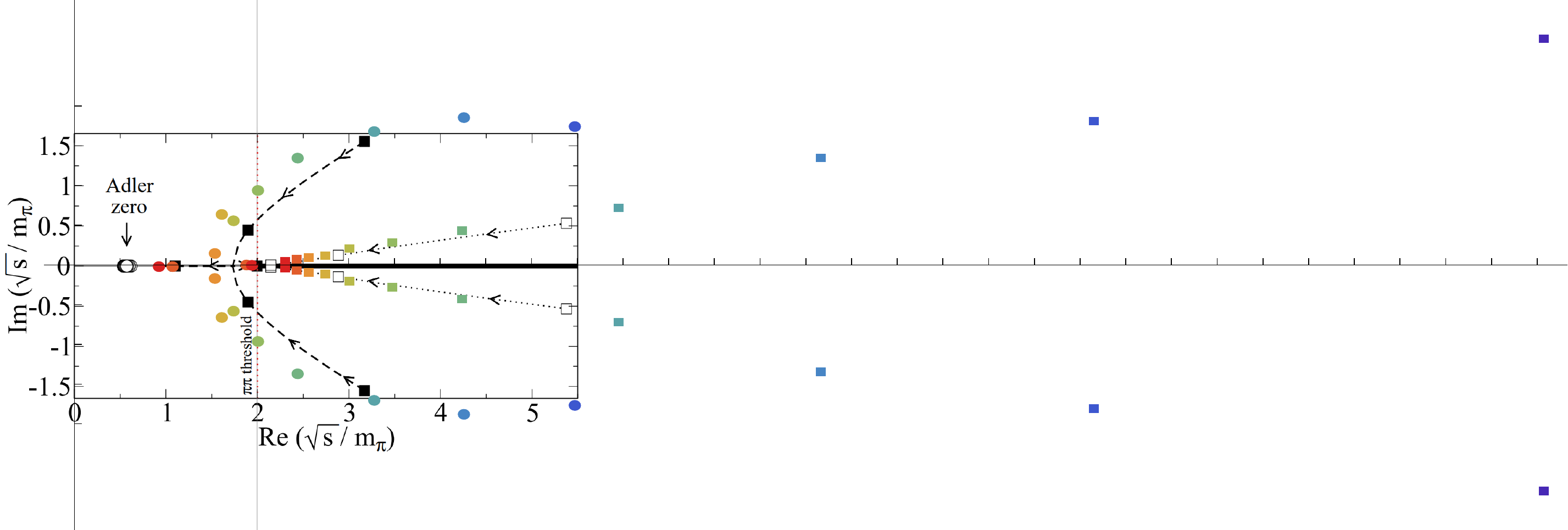}
	\caption{The $\rho$ and $\sigma$ meson poles in the complex plane of energy in units of the pions mass $m_\pi$. The coloured points are our numerical study with varying quark mass. The black overlay is taken from \cite{Hanhart:2008mx} showing the same from a unitarized chiral perturbation theory analysis.}
	\label{pelaez}
\end{figure}

As a taster of the physics we can observe using GTB, we provide a first overview of results for the $\rho$ and $\sigma$ mesons in Figure \ref{pelaez}. Here we show the position of the $\rho$ and $\sigma$ poles in the complex plane of energy. Our points, in colour, are for a particular set of parameters in the analysis and we do not here include the error bars we compute more carefully below. Omitting them allows us to more clearly see the base physics  and allows us to overlay the results of Pelaez et al \cite{Hanhart:2008mx} (in black) who have extracted a similar plot purely using low energy pion data. Here we are plotting in units of the pion mass so small quark mass ({\color{blue}{bluer}}) lies to the right  and large quark mass ({\color{red}{redder}}) to the left. The inner V-shape are for the $\rho$ meson mass (real part) and width (imaginary part). At large quark mass the pions become heavy and the decay channel of the $\rho$ falls and hence the width reduces to zero. Our analysis lines (coloured points) agree well to the lines of \cite{Hanhart:2008mx} (black). The outer set of points are for the $\sigma$ meson and also show the shutting off of the decay channel to pions as the pion mass rises. Here we can track the width to zero when the $\sigma$ poles coming from positive and negative imaginary pieces meet and separate along the real axis. 
Our results are a little shifted from the chiral perturbation theory results of \cite{Hanhart:2008mx} but the expected structure is all present. Below we will present a more careful analysis of these states including estimates of errors. 

The paper is organized as following. After a brief overview of the GTB approach in the next section, we describe how to change the quark mass/pion mass in the GTB setup in section \ref{dial}.  Section \ref{numpara} describes the numerical parameters and error estimation we use in this paper. In section \ref{specresu}, we present the results on the spectra of $\sigma,\rho,f_2$ for varying quark/pion mass. These results are summarized in section \ref{sum}. In section \ref{lec}, we analyze the $m_\pi$ dependence of the low energy parameters (the Gasser-Leutwyler coefficients of the chiral Lagrangian, pion radii etc). We finish with the conclusions. Additional details about the numerical implementation are given in the appendix. The numerical program for generating the results of the paper can be found in the github repository \href{https://github.com/hyfysics/gauge-theory-bootstrap}{gauge-theory-bootstrap}.

\section{Overview of the Gauge Theory Bootstrap}

We start by briefly reviewing the Gauge Theory Bootstrap (GTB) approach \cite{PhysRevLett.133.191601,PhysRevD.110.096001,He:2024nwd,He:2025gws}. The unitarity of the theory is encoded by a positive semidefinite condition on the matrix of state overlaps including pion states and states created by local operators. The matrix elements are then given by the scattering amplitudes, form factors and current correlators. At the order we work, we consider only 2-pion states and one operator for a given set of quantum numbers. The unitarity constraint reads \cite{Karateev:2019ymz}:
\beq\label{Bmt}
 \begin{array}{c c} &
   	\begin{array}{c c c} \ket{ \pi \pi\mbox{(in)}}_{P,I,\ell} & \ket{ \pi \pi\mbox{(out)}}_{P,I,\ell} & \cO_{P,I,\ell}\ket{0} \\
   	\end{array}
   	\\
   	\begin{array}{c c c}
   		\bra{ \pi \pi\mbox{(in)}}_{P',I,\ell}  \\
   		\bra{ \pi \pi\mbox{(out)}}_{P',I,\ell}\\
   		\bra{0} \cO^\dagger_{P',I,\ell}
   	\end{array}
   	&
   	\left(
   	\begin{array}{c c c}
   		1 &\ S^I_\ell(s)\ &\ \ \cF^I_\ell(s) \\
   		S_\ell^{I*}(s) &\ 1\ & \ \ \cF^{I*}_\ell(s) \\
   		\cF^{I*}_\ell(s) &\ \cF^I_\ell(s)\ &\ \ \rho^I_\ell(s)
   	\end{array}
   	\right) 
   \end{array} \succeq 0,\;\; \forall I,\ell, \ s>4m^2_{\pi},\;\; 
\eeq
   where we removed an overall factor $\delta^4(P-P')$ that ensures momentum conservation. The $S^I_\ell(s),\mathcal{F}^I_\ell(s),\rho^I_\ell(s)$ are the 2-to-2 partial amplitude, 2-pion form factor and spectral density for a given center-of-mass energy $\sqrt{s}$, isospin $I$ and angular momentum $\ell$.
We will now work through these quantities in turn.

\subsection{2-to-2 scattering amplitude $S_\ell^I$}\label{SIl}

In terms of the standard Mandelstam variables $s,t,u$, we write the 2-to-2 amplitude (splitting it by isospin channel) as
\begin{equation} \begin{array}{ccl}
    T_{ab,cd} &= &A(s,t,u) \delta_{ab} \delta_{cd}+ A(t,s,u) \delta_{ac}\delta_{bd} + A(u,t,s) \delta_{ad}\delta_{bc}\\
    &&\\
    &=& {1 \over 3}T^{I=0} \delta_{ab} \delta_{cd} + {1 \over 2} T^{I=1} ( \delta_{ac}\delta_{bd} - \delta_{ad} \delta_{bc}) +{1 \over 2} T^{I=2}(\delta_{ac}\delta_{bd}+\delta_{ad}\delta_{bc} -{2 \over 3} \delta_{ab}\delta_{cd})
    \end{array}
\end{equation}
where the amplitude $A(s,t,u)$ is an analytic function in $s,t,u$ with cuts for $s,t,u\ge 4m_\pi^2$, satisfying $A(s,t,u)=A(s,u,t)$ and we have the on-shell condition $s+t+u = 4 m_\pi^2$. The amplitude can be parameterized in a double dispersion form as
\begin{equation} \begin{array}{ccl}
    A(s,t,u)  &=& T_0 +{1 \over \pi} \int_{4m_\pi^2}^\infty dx\, {\sigma_1(x) \over x-s} + {1 \over \pi}  \int_{4m_\pi^2}^\infty dx \, \sigma_2(x) \left[ {1\over x-t} + {1 \over x-u} \right]\\
    &&\\ &&+ {1 \over \pi^2} \int_{4 m_\pi^2}^\infty dx \int_{4 m_\pi^2}^\infty  dy \left[{\rho_1(x,y) \over (x-s)(y-t)} + {\rho_1(x,y) \over (x-s)(y-u)} + {\rho_2(x,y) \over (x-t)(y-u)} \right]
    \end{array}\label{amplitude}
\end{equation}
\normalsize
which has the expected analyticity and crossing properties. The parameters are then $\{T_0, \sigma_1(x),\sigma_2(x),\rho_1(x,y), \rho_2(x,y)\}$.

From the amplitudes, we can project out the partial waves with angular momentum $\ell$ in each isospin channel $I$ as
\begin{equation}
    f_\ell^I(s) = {1 \over 4} \int_{-1}^1 d \mu P_\ell(\mu)T^I(s,t),\hspace{0.75cm}t=\frac{(4m_\pi^2-s)(1-\mu)}{2}
\end{equation}
with $P_\ell$ the Legendre polynomials. $I,\ell$ should be both even or both odd.
The $S^I_\ell$  in (\ref{Bmt}) is then given by
\begin{equation}
    S^I_\ell(s) = 1 + i \pi \sqrt{s-4 m_\pi^2 \over s} f^I_\ell(s^+),\;\; s\in \mathbb{R}_{\ge 4 m_{\pi}^2}
\end{equation}
In general, $S^I_\ell(s)$ satisfies the partial waves unitarity
\begin{equation}\label{unit}
    |S^I_\ell(s)|\le 1,\;\; s>4m_\pi^2
\end{equation}
which is a condition from the positive semi-definiteness on the upper-left 2-by-2 corner of \eqref{Bmt}.

At very low energy, the amplitude is well described by the tree-level amplitude of the nonlinear sigma model, i.e. the Weinberg model:
\begin{equation}\label{treeA}
    A_{\text{tree}}(s,t,u) = {4 \over \pi}{ s-m_\pi^2 \over 32 \pi f_\pi^2}
\end{equation}
whose partial waves are given by
\begin{equation}
\begin{array}{l}
    f^0_{0,\text{tree}}(s)= {2 \over \pi}{2s - m_\pi^2 \over 32 \pi f_\pi^2}, ~~~ f^1_{1,\text{tree}}(s)= {2 \over \pi}{s - 4 m_\pi^2 \over 96 \pi f_\pi^2}, \\  \\f^0_{2,\text{tree}}(s)=0, ~~~
    f^2_{0,\text{tree}}(s)= {2 \over \pi}{2 m_\pi^2-s  \over 32 \pi f_\pi^2}
    \end{array}
\end{equation}
In implementing GTB, we require the matching to these amplitudes at a set of unphysical energies \cite{PhysRevD.110.096001}:
\begin{equation} \label{firstchi}
\begin{aligned}
    &||[f^0_0(s_j)-R^\text{tree}_{01}(s_j)f^1_1(s_j),f^2_0(s_j)-R^\text{tree}_{21}(s_j)f^1_1(s_j),f^0_2(s_j)]||\le \epsilon^\chi\\
    &R^\text{tree}_{01}=\frac{f^0_{0,\text{tree}}}{f^1_{1,\text{tree}}},\;R^\text{tree}_{21}=\frac{f^2_{0,\text{tree}}}{f^1_{1,\text{tree}}},\;s_j=\{1/2,1,3/2,2\}
\end{aligned}
\end{equation}

The UV behavior of the scattering amplitude remains unconstrained since we cannot compute pion scattering from QCD. In fact at high energies we expect the scattering to be inelastic such that $S^I_\ell(s\rightarrow\infty)\rightarrow 0$. \footnote{Similarly at high energies we have $\rho\gg |\cF|^2$ for the form factor and spectral density described below.} In that case positivity of the matrix in \eqref{Bmt} is trivial. 

\smallskip

To connect with the UV theory of QCD we have to introduce local QCD operators and describe their form factors and spectral densities. 

\subsection{Form factor $F_\ell^I$} \label{ff}

Following \cite{He:2024nwd,He:2025gws}, we focus on the $S0,P1,D0$ channels and consider the QCD operators of scalar, vector currents and the energy-momentum tensor:\footnote{$\Delta=\frac{1}{\sqrt{2}}(0,1,i,0)$}
\begin{subequations}\label{currents}
	\beqa
	j_S &\equiv& m_q(\bar{u}u+\bar{d}d) = \frac{1}{2}m^2_{\pi}\pi^a\pi^a +\cO(\pi^4) \label{jscalar}\\
	j_V &\equiv& \frac{1}{2}(\bar{u}\gamma^\mu u-\bar{d}\gamma^\mu d)\Delta_\mu  = \epsilon^{abc} \pi^b \partial_\Delta \pi^c +\cO(\pi^4) \\
	j_T &\equiv& T^{\mu\nu}\Delta_\mu\Delta_\nu
	\eeqa
\end{subequations}
Their 2-pion form factor is given by the overlap of a state created by the operator with the 2-pion state:
\begin{equation}
{}_{\text{out}}\langle\pi^-(p_1)\pi^+(p_2)|\mathcal{O}^I_\ell(0)|0\rangle=F^I_\ell(s),\;\; s=(p_1+p_2)^2
\end{equation}
and in the basis of the 2-pion state with definite isospin and angular momentum, we have the rescaled form factor $\mathcal{F}^I_\ell$:
\begin{equation}
{}_{\text{out}}\langle P,I,\ell|\mathcal{O}^I_\ell(0)|0\rangle=\mathcal{F}^I_\ell(s),\;\; s=P^2
\end{equation}
where the two quantities are related by:
\begin{subequations}\label{FcurlyF}
    \begin{eqnarray}
        \mathcal{F}_0^0(s) &= \frac{\sqrt{6\pi}}{16\pi^3} \frac{1}{s^{\frac{1}{4}}} \left(\frac{s-4m_\pi^2}{4}\right)^{\frac{1}{4}} F_0^0(s) \\
    \mathcal{F}_1^1(s) &= \sqrt{\frac{4\pi}{3}} \frac{1}{8\pi^3} \frac{1}{s^{\frac{1}{4}}} \left(\frac{s-4m_\pi^2}{4}\right)^{\frac{3}{4}} F_1^1(s) \\
    {\cal F}^0_2(s) &= \frac{1}{4\pi^3} \sqrt{\frac{\pi}{5}}\frac{1}{s^{\frac{1}{4}}}\left(\frac{s-4m_\pi^2}{4}\right)^{\frac{5}{4}}F^{0}_2(s)
\end{eqnarray}
\end{subequations}

The form factors $F^I_\ell(s)$ are analytic functions of $s$ with cuts for $s>4m_\pi^2$ and can be parameterized using a dispersion relation,
\begin{equation}
    F_\ell^I(s) = F^I_\ell(0) + {s \over \pi} \int_{4 m_\pi^2}^\infty dx {\text{Im} F^I_\ell(x) \over x(x-s)},
\end{equation}
by the imaginary part above the cut. Due to the symmetry constrained forms of the currents \eqref{currents}, their normalization at zero energy is fixed: $F^1_1(0)=1$, $F_0^0(0)=m_\pi^2$, $F^0_2(0)=1$.

In the UV, the form factors approach the Lepage-Brodsky form \cite{osti_1447331}. Since we will constrain them at only high (but not asymptotically high) energies, we allow a coefficient in each case $\chi^I_\ell$ \cite{He:2025gws} and require the form factors to be bounded above energies $s_0\simeq 2\text{GeV}$:
\begin{subequations}\label{chiIL}
\begin{eqnarray}
      |F_0^0(s>s_0)| &\le& \chi^0_0 {m_\pi^4 \over s} \ln \left( {s \over m_\pi^2} \right) \\
  |F_{1}^1(s>s_0)| &\le& \chi_1^1 {16 \pi \alpha_s f_\pi^2 \over s}\\
    |F_2^0(s>s_0)| &\le& \chi_2^0 { 48 \pi \alpha_sf_\pi^2 \over s}
\end{eqnarray}
\end{subequations}
Here $f_\pi$ enters as an estimate of the size of the pion and $\alpha_s$ is the strong coupling evaluated at $s_0$.

\subsection{Current correlator $\rho_\ell^I$}

Further UV information comes from considering the 2-point current correlator
\begin{equation}
    \langle0|\mathcal{O}^{I\dagger}_\ell(P)\mathcal{O}^{I}_\ell(P')|0\rangle=\rho^I_\ell(s)\delta^{(4)}(P-P')
\end{equation}
which is related to the time-ordered 2-point function 
\begin{equation}
    \Pi^I_\ell(s)=i\int\frac{d^4 x}{(2\pi)^4}e^{ip\cdot x}\langle 0|\hat{T}\{\mathcal{O}^{I\dagger}_\ell(x)\mathcal{O}^{I}_\ell(0)\}|0\rangle
\end{equation}
by
\begin{equation}
    2\text{Im}\Pi^I_\ell(s+i\epsilon)=\rho^I_\ell(s)
\end{equation}
In the UV, the $\Pi^I_\ell$ are given by perturbative QCD computations which for example take the following form for the currents in \eqref{currents}  (for $N_f=2,N_c=3$) \cite{Chetyrkin:1996sr,Chetyrkin:1996ez,Gorishnii:1990vf,Zoller:2012qv}: 
\begin{equation}
	\begin{aligned}
		&\Pi^0_0(s) = -\frac{m_q^2}{(2\pi)^4}\frac{s}{8\pi^2}  \Bigg\{ 6\ln(-s/\mu^2) +\frac{\alpha_s}{\pi} \left[ 34\ln (-s/\mu^2)- 6 \ln^2(-s/\mu^2)\right]  \\
		&  +\left(\frac{\alpha_s}{\pi}\right)^2 
		\left[\left(\frac{10021}{24}-109\zeta(3)\right)\ln (-s/\mu^2)-\frac{296}{3} \ln^2 (-s/\mu^2)+\frac{53}{6}\ln^3 (-s/\mu^2)\right]+O(\alpha_s^3)\Bigg\}\label{Pi00}
	\end{aligned}
\end{equation}
\begin{equation}
	\begin{aligned}
		\Pi^1_1 &=  -\frac{1}{(2\pi)^4}\frac{s}{8\pi^2} \Bigg\{\ln(-s/\mu^2) +\frac{\alpha_s}{\pi} \ln(-s/\mu^2)  \\
		&  +\left(\frac{\alpha_s}{\pi}\right)^2 
		\left[\left(\frac{107}{8}-\frac{29\zeta(3)}{3}\right)\ln (-s/\mu^2)-\frac{29}{24} \ln^2 (-s/\mu^2)\right]+O(\alpha_s^3) \Bigg\}\label{Pi11}
	\end{aligned}
\end{equation}
\begin{equation}
	\Pi^0_2(s)=-\frac{1}{(2\pi)^4}\frac{s^2}{8\pi^2}\left(\frac{11}{10}\ln (-s/\mu^2)-\frac{17}{18}\frac{\alpha_s}{\pi}\ln (-s/\mu^2)+O(\alpha_s^2)\right)\label{Pi02}
\end{equation}
The UV information puts constraints on the moments of the spectral density $\rho^I_\ell$ through the finite energy sum rules \cite{PhysRevLett.133.191601,PhysRevD.110.096001}:
\begin{align}
    \int_{4m_\pi^2}^{s_0} \rho(x) x^n dx = -s_0^{n+1} \int_{0}^{2\pi} e^{i(n+1)\varphi} \Pi(s_0 e^{i\varphi}) d\varphi\label{sr}
\end{align}
where the $\Pi(s_0 e^{i\varphi})$ is evaluated at large energy scale $s_0\simeq 2\text{GeV}$.

Furthermore, in the IR, the 2-point function is dominated by the free pion kinematics \cite{GASSER1984142}:
\begin{subequations}\label{rholow}
\begin{eqnarray}
        \rho^0_0 &\simeq& {m_\pi^4 \over (2 \pi)^4} {3 \over 16 \pi} \left( 1 - {4 m_\pi^2 \over s} \right)^{1/2}\\
    \rho^1_1 &\simeq& {1 \over (2 \pi)^4} {s \over 24 \pi} \left( 1 - {4 m_\pi^2 \over s} \right)^{3/2}\\
    \rho^0_2 &\simeq& {1 \over (2 \pi)^4} {s^2 \over 160 \pi} \left( 1 - {4 m_\pi^2 \over s} \right)^{5/2}
\end{eqnarray}
\end{subequations}
Note that this is the square of the rescaling factor in \eqref{FcurlyF}, reflecting the fact that at low energies $\rho=|\mathcal{F}|^2$ and 2-pion states dominate.

\subsection{Watsonian iteration}

The set of conditions described above from analyticity, crossing, unitarity and dynamical UV information from QCD constrain the bootstrap variables 
\begin{equation}\label{para}
    \{T_0, \sigma_1,\sigma_2,\rho_1, \rho_2,\text{Im} F^I_\ell, \rho_\ell^I\}
\end{equation} 
As demonstrated in previous work \cite{He:2025gws}, the resulting space of solutions -- as characterized by a set of low energy parameters -- is highly constrained and these solutions display qualitatively similar low energy dynamics. Furthermore, a certain solution can be identified by iteratively searching for the saturation of the matrix \eqref{Bmt} based on the assumption that in the intermediate energies the physics is dominated by the 2-pion states. The iterating procedure is dubbed Watsonian iteration in \cite{He:2025gws}.

In practice, as the initial step (i.e., 0th iteration), one starts with an arbitrary initial functional which is  maximized. This finds an initial set of solutions for the parameters \eqref{para}. Subsequently, for each iteration step, based on such a solution (referred to as ``old" below), one can construct a linear functional in the bootstrap variables (referred to as ``new" below) to maximize and obtain a new solution. Iterating the procedure until the results converge yields the GTB solution. Such linear iterating functionals can be taken to be
\begin{equation}
    \begin{aligned}
       \mathcal{F}_1=\int ds\sum_{I,\ell}\left\{\frac{|h^I_\ell|}{|F^I_\ell|}\bigg|_{\text{old}}(\text{Re}F^I_{\ell}|_{\text{old}}\text{Re}h^I_{\ell}|_{\text{new}}+\text{Im}F^I_{\ell}|_{\text{old}}\text{Im}h^I_{\ell}|_{\text{new}})-\text{Im}h^I_{\ell}|_{\text{new}}\right\}
    \end{aligned}
\end{equation}
and
\begin{equation}
    \mathcal{F}_2=\int ds\sum_{I,\ell}\left\{2\text{Re}\mathcal{F}^I_{\ell}|_{\text{old}}\text{Re}\mathcal{F}^I_{\ell}|_{\text{new}}+2\text{Im}\mathcal{F}^I_{\ell}|_{\text{old}}\text{Im}\mathcal{F}^I_{\ell}|_{\text{new}}-\rho^I_\ell|_{\text{new}}\right\}
\end{equation}
Here, the $\mathcal{F}_1$ comes from linearizing the unitarity constraints \eqref{unit} and applying the Watson theorem $S=\frac{F}{F^*}$ and its maximization pushes for unitarity saturation while aligning the phases of the form factor and the amplitude. The functional $\mathcal{F}_2$ comes from linearizing the condition
\begin{equation}
    |\mathcal{F}^I_\ell|^2\le \rho^I_\ell
\end{equation}
from \eqref{Bmt} and maximizing it pushes for saturation. Together, these two functionals push for saturation of the positive semidefinite matrix \eqref{Bmt}. See section 2.2 of \cite{He:2025gws} for more details. 

As tested in \cite{He:2025gws}, the initial functional can be chosen by finding feasible points of the set of constraints by minimizing 0. Denoting the iteration step counter as $n_W$, the maximization functional is:
\begin{equation}
\text{max}\;\mathcal{F}_{n_W}=\begin{cases}
0,&n_W=0\\
\mathcal{F}_1+\mathcal{F}_2|_{\text{old}\to n_W-1},&n_W\ge 1
\end{cases}
\end{equation}
where $|_{\text{old}\to n_W-1}$ means the functional is constructed from the results of the $n_W-1$-th iteration. Iterating until convergence allows us to identify the solution shown in this paper.


\section{Dialing the quark mass $m_q$}\label{dial}

The purpose of this work is to understand the dependence of the strong QCD dynamics -- meson spectrum ($\sigma$, $\rho$, $f_2$) and low energy parameters -- on the quark mass, $m_q$. We assume the $u,d$ quarks are degenerate here. As clear from the overview in the previous section, the quark mass appears explicitly in the UV sum rule for the scalar current:
\begin{equation}
	j_S \equiv m_q(\bar{u}u+\bar{d}d) = \frac{1}{2}m^2_{\pi}\pi^a\pi^a +\cO(\pi^4) \label{jscalar}
\end{equation}
whose two point function is given by \eqref{Pi00} which at 2-loop is:
\begin{equation}
	\begin{aligned}
		\Pi^0_0(s) = -\frac{m_q^2}{(2\pi)^4}\frac{ s}{8\pi^2} \Bigg\{ 6\ln(-s/\mu^2) +\frac{\alpha_s}{\pi} \left[ 34\ln (-s/\mu^2)- 6 \ln^2(-s/\mu^2)\right] +O(\alpha_s^2)\Bigg\}
	\end{aligned}
\end{equation}
leading to the sum rule:
\begin{equation}
	\begin{aligned}
		&\frac{1}{s_0^{n+2}}\int_{4 m_{\pi}^2}^{s_0} \rho^0_0(x) x^n dx = \frac{m_q^2}{(2\pi)^4} \frac{3}{2\pi(n+2)} \Bigg\{1+\frac{\alpha_s}{\pi} \left[\frac{17}{3} + \frac{2}{n+2}\right] +O(\alpha_s^2)\Bigg\}
	\end{aligned}
\end{equation}
So the quark mass is a free parameter to change in the UV input through the sum rules.

In addition, the quark mass $m_q$ also appears implicitly through the pion mass $m_\pi$ by the Gell-Mann Oakes-Renner  (GMOR) relation \cite{Gell-Mann:1968hlm}: 
\begin{equation}
	m_{\pi}^2=-\frac{m_q}{f_{\pi}^2}\langle0|\bar{q}q|0\rangle
\end{equation}
In this work, we consider only the leading order and assume the quark condensate and value of $f_\pi$ do not change. We thus change $\mpi$ in conjunction with $m_q$. Note that in the numerics, we use the pion mass to set the energy unit for the dimensionful quantities in the computations ($s_0,f_\pi,m_q$) where we take $m_\pi=1$.

In principle, the range of values that we can explore is $0\le\mpi\le\text{min}\{\half m_\rho, \half m_\sigma\}$. The upper limit is the point where the $\sigma$ or $\rho$ meson becomes stable and should be include as an asymptotic state requiring a modification of the GTB setup. Notice that as we increase the pion mass, the masses of the $\sigma,\rho$ mesons also increase so we cannot determine a priori this range, but later results in this paper give a value $\mpi\simeq 492\,\MeV$ for when the $\rho$ resonance becomes stable. 
For the computations whose results will be presented in the next section we choose twelve values:
\begin{equation}\label{mqlist}
	m_q = \left\{\frac{1}{8},\frac{1}{4},\frac{1}{2},1,2,3,4,5,6,7,8,9\right\}\times m^{\text{phys}}_q
\end{equation}
and 
\begin{equation}\label{mpilist}
	m_\pi = \left\{\frac{1}{2\sqrt{2}},\frac{1}{2},\frac{1}{\sqrt{2}},1,\sqrt{2},\sqrt{3},2,\sqrt{5},\sqrt{6},\sqrt{7},2\sqrt{2},3\right\}\times m^{\text{phys}}_\pi
\end{equation}
where the physical quark mass and pion mass are taken to be
\begin{equation}
	m^{\text{phys}}_q(\text{2 GeV}) = 5.25\,\MeV, \ \ \ m^{\text{phys}}_\pi=140\,\MeV
\end{equation}
This corresponds to the range of pion mass
\begin{equation}
   50\,\MeV \lesssim \mpi \lesssim 420\,\MeV 
\end{equation}
where the results near the end of the range are not always feasible and depend on the numerical parameters (see below).

\section{Numerical parameters and error estimation}\label{numpara}

There are several numerical factors that we adjust in our analysis. These appear in eq. \eqref{firstchi} and eq. \eqref{chiIL} and we discuss their role here.

The factor $\epsilon^\chi$ in eq. \eqref{firstchi} enforces the matching of the partial waves with the tree-level results at very low energies. The reference value of $\epsilon^{\chi}=0.003$ for physical pion mass $m_\pi=m^{\text{phys}}_\pi$ results from the initial GTB exploration \cite{PhysRevLett.133.191601,PhysRevD.110.096001} and as demonstrated in \cite{He:2025gws}, there is a range of $\epsilon^\chi$ where the converged results are not affected. As we change the pion mass, the matching error $\epsilon^\chi$ should change accordingly since the expression \eqref{firstchi} approximately captures the deviation from the tree-level prediction and according to $\chi$PT \cite{GASSER1984142}, this increases with the pion mass.
For any given choice of $m_q$ we reduce $\epsilon^\chi$ to near the minimum value where the numerical program is still feasible. The set of $\epsilon^{\chi}$ we use corresponding to the set of $m_q$ values in eq. \eqref{mqlist} are
\begin{equation}
    \epsilon^{\chi}=\left\{\frac{1}{2500},\frac{1}{1250},0.0016,\frac{3}{1000},\frac{9}{1000},\frac{1}{50},\frac{1}{25},\frac{1}{20},\frac{2}{25},0.12,0.16,\frac{2}{5}\right\}
\end{equation}

As one varies the quark mass $m_q$, the Brodsky-Lepage asymptotic form factors do not change. However, as described in section \ref{ff}, a numerical multiplier $\chi^I_\ell$ is introduced to account for the form factor matching with the Brodsky-Lepage form at high (but not asymptotically high) energies. As in \cite{He:2025gws}, we choose the near-minimal values of $\chi^I_\ell$ where a feasible solution exists.  We find these base values: 
\begin{equation} \label{base}
\begin{aligned}
    &\chi^0_0=4,\;\chi^1_1=1.5,\; &\chi^0_2=4
\end{aligned}
\end{equation}
The choices of these $\chi$ parameters is clearly not fixed from first principles and so the dependence upon them provides a measure of the errors in the method. To provide some estimate of this error we have also run the code with the $\chi^I_\ell$'s doubled. Our full set of choices are 
\begin{equation}\label{chival}
\begin{aligned}
    &\chi^0_0=\{4,8\},\;\chi^1_1=\{1.5,3\},\; &\chi^0_2=\{4,8\}
\end{aligned}
\end{equation}
generating 8 total set of results. To make error estimates, for each quantity we compute below, we find the average and standard deviation from these 8 sets of results. Whilst imperfect this at least provides a sensible understanding of the typical errors.

Finally we note that for the sum rules, we use the 3-loop results for $S0,P1$ \cite{He:2025gws} and 2-loop results for $D0$ \cite{Zoller:2012qv} and set the sum rules to be equal to the perturbative QCD values with a $5\%$ error for each moment. This takes care of the fact that the $S0$ sum rules depends on $m_q^2$ and the error should also depend on it proportionally.

\section{Spectrum results}\label{specresu}

Currently, the GTB computations of strongly coupled resonances concentrate on three main channels, $S0$, $P1$ and $D0$. The dominant resonances in each of those channels are the $\sigma$, $\rho$ and $f_2$ mesons. The main result of this paper is the pion mass dependence of those three resonances. We describe each of these channels starting with the $P1$ channel since the $\rho$ properties are most closely determined. 

\subsection{The $\rho$ meson, $P1$ channel}

\begin{figure}[t]
	\centering
	\includegraphics[width=\textwidth]{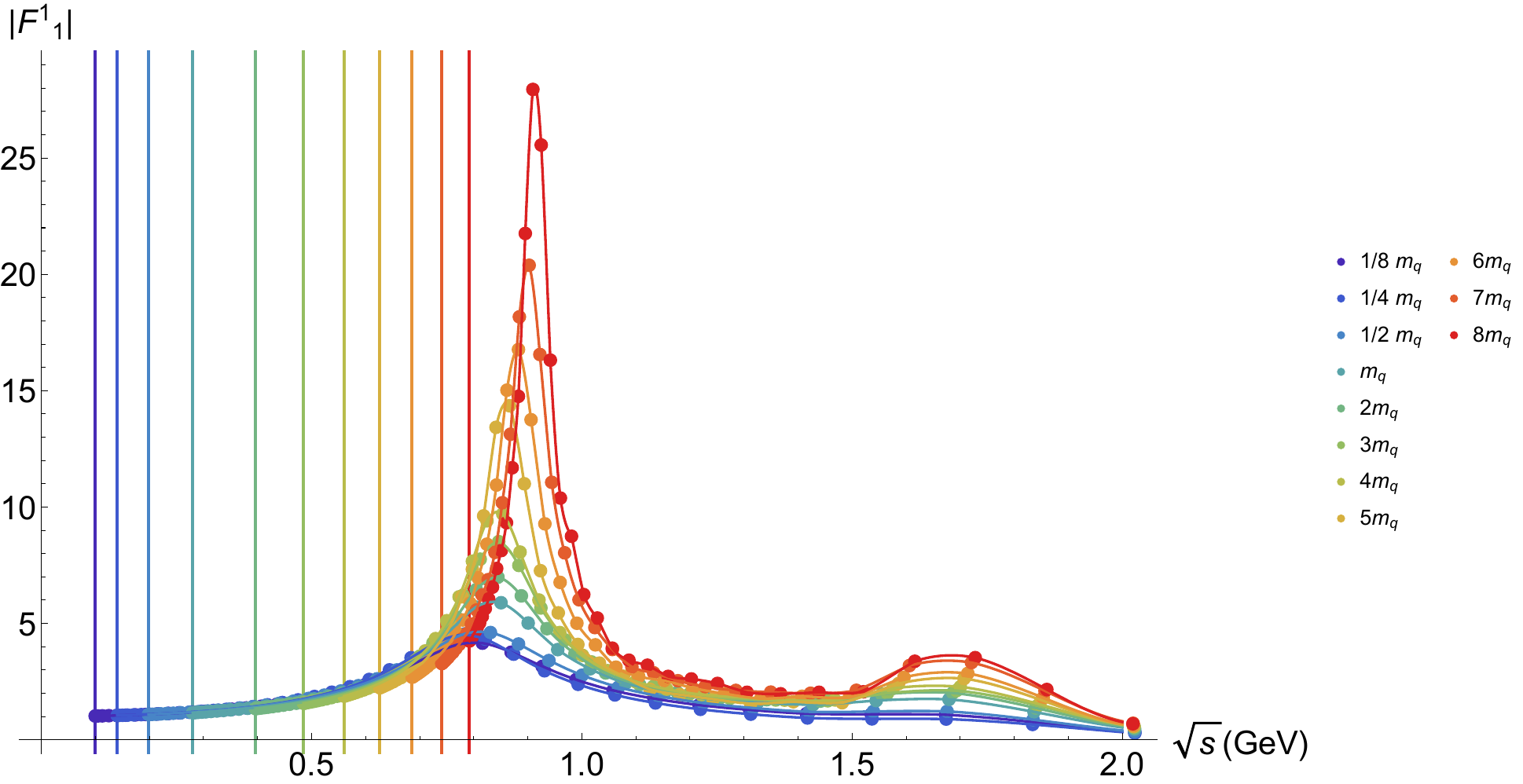}
	\caption{Modulus of vector form factor $|F^1_1|$ as a function of energy $\sqrt{s}$ for varying quark mass. The gridlines with corresponding color indicates the moving threshold $\sqrt{s}=2m_{\pi}$. The legend corresponds to the range of $m_q$ values \eqref{mqlist} where we neglect the superscript $^{\text{phys}}$.}
	\label{FFP1}
\end{figure}

The $P1$ channel corresponds to isospin $I=1$ and angular momentum $\ell=1$. 
In fig. \ref{FFP1}, we show the modulus of the form factor as a function of the energy $\sqrt{s}$ in physical unit. The form factor has a clear, narrow $\rho$ resonance peak for all the quark/pion masses that we considered. As the mass of the pion increases, the center of the peak slightly increases and the width of the peak decreases. Note that the threshold $2m_\pi$ increases with pion mass and we indicate this with vertical lines of the same color in fig. \ref{FFP1}. Overall, the $\rho$ peak moves closer to the threshold with increasing $m_\pi$. Physically, as the pion mass increases, there is less phase space available for the decay and the resonance becomes narrower. Eventually it would become stable and the $\rho$ meson should be included as an asymptotic particle changing the basic bootstrap setup. We do not consider that case in this paper.

\begin{figure}[t]
	\centering
	\includegraphics[width=0.9\textwidth]{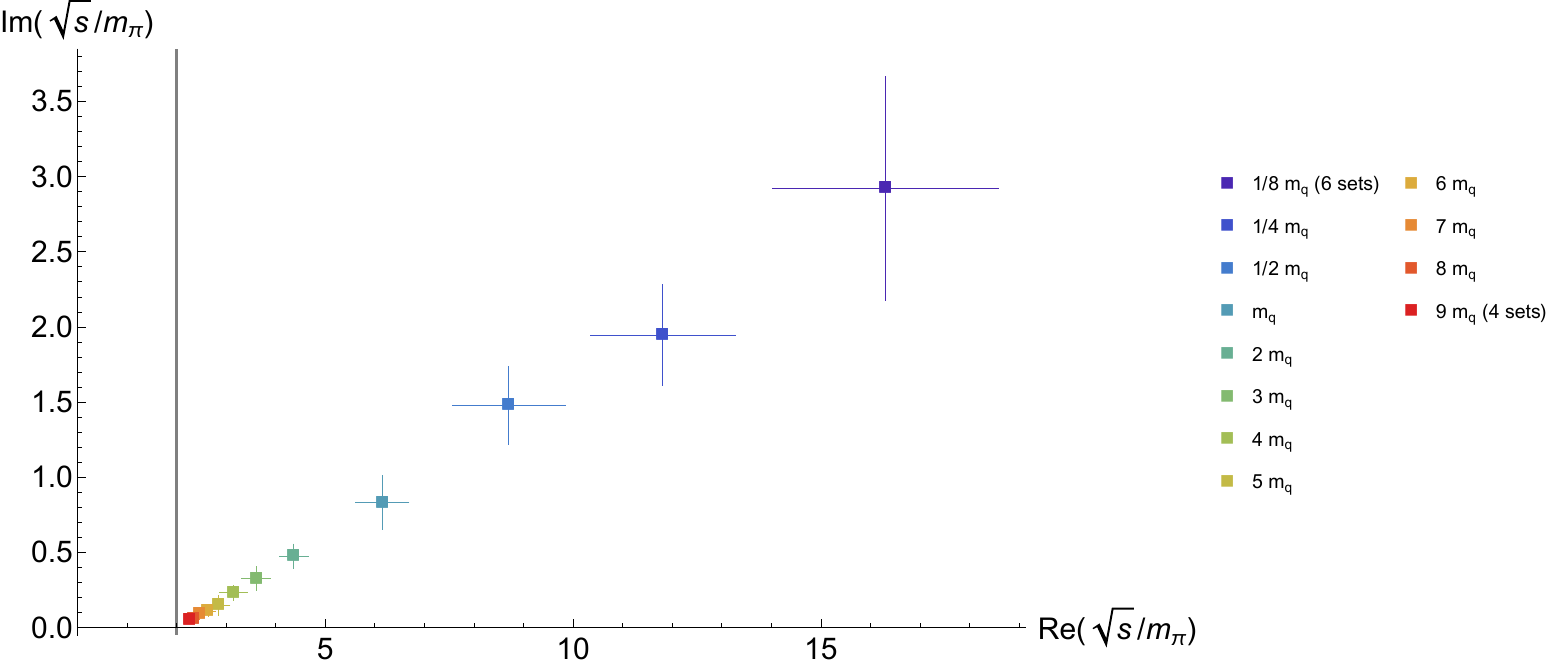}
	\caption{The location of the $\rho$ meson pole in the complex plane of $\sqrt{s}$ for varying quark mass where $\sqrt{s_{\text{pole}}}=m_\rho+i\Gamma_\rho/2$. The plot is done in units of the pion mass where $m_\pi=1$. The results and $2\sigma$ error bars for each quark mass are obtained by averaging over the set \eqref{chival} where for the two ends values of the quark mass range \eqref{mqlist}, only a subset of the \eqref{chival} are feasible which we average over. The legend corresponds to the range of $m_q$ values \eqref{mqlist} where we neglect the superscript $^{\text{phys}}$.}
	\label{rhopole}
\end{figure}

From the bootstrap results, we obtain the full set of amplitude parameters \\ $\{T_0, \sigma_{1,2},\rho_{1,2}\}$ which, from the description in section \ref{SIl}, can be used to evaluate the analytic partial amplitude
\begin{equation}
    S^I_\ell(s)=1+\pi\sqrt{\frac{4m_\pi^2-s}{s}}f^I_\ell(s),\;\; s\in \mathbb{C}-(-\infty,0)\cap(4,\infty)
\end{equation}
at any complex $s$ point on the first sheet. 
\begin{figure}[h!]
\centering
\includegraphics[width=0.8\textwidth]{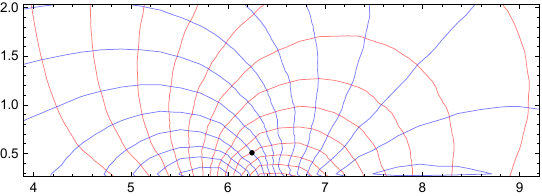}
\caption{An illustration of the contour lines of $\text{Re}S^I_\ell$ ({\color{red}red}) and $\text{Im}S^I_\ell$ ({\color{blue}blue}). The intersection of their zero contour line is labeled with the black dot which is the zero of $S^I_\ell(s)$.}
\label{rhozero}
\end{figure}
By unitarity saturation (enforced by the Watsonian iteration procedure), the resonance pole on the second Riemann sheet of $S^I_\ell(s)$ gives rise to a zero on the first sheet and we can extract the position of the zeros by evaluating $S^I_\ell(s)$ on the first sheet. In practice, we evaluate the $S^I_\ell(s)$ on a grid of complex $s$ points to find the contour lines of $\text{Re}S$ and $\text{Im}S$. The zero can be extracted from the coordinate of the intersection of the zero contour line $\text{Re}S$ and $\text{Im}S$. See fig. \ref{rhozero} for an illustration.
We repeat this computation across the range of masses in \eqref{mqlist} and, for each mass, we run the computation for the choices of $\chi$ parameters in eq. \eqref{chival} to extract the locations of the zero/pole and compute the mean and standard deviation from the 8 set of results \eqref{chival}. Note that for the two ends values of the quark mass range \eqref{mqlist}, not all 8 sets of \eqref{chival} give feasible results. So for these results, we only average over 4 and 6 sets of feasible results. (This applies also to the computations of other quantities below.) In fig. \ref{rhopole}, we plot the mean locations and $2\sigma$ error bars of the $\rho$ meson pole on the complex plane $\sqrt{s_\text{pole}}$ for varying quark mass. From the real and imaginary parts of the pole location, we can extract the mass and half-width of the $\rho$ meson:
\begin{equation}
    \sqrt{s_\text{pole}}=m+i\Gamma/2
\end{equation}\label{polemG}
Note here we are plotting in units of the pion mass, namely that for each pion mass, the value $\sqrt{s_\text{pole}}$ is given by setting $m_\pi=1$. This explains why as the pion mass falls (more blue) the error bars look larger.

\begin{figure}[h!]
	\centering	\includegraphics[width=0.75\textwidth]{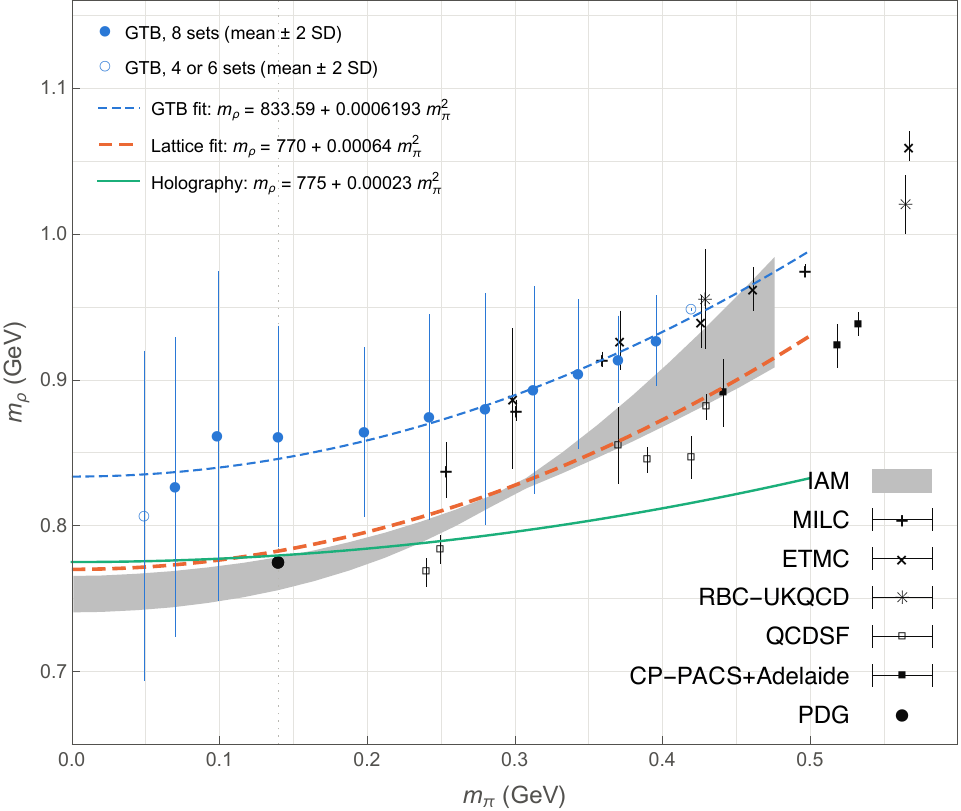}
	\caption{Comparison of the GTB results with lattice, holographic and unitarized $\chi$PT results. The GTB computations are shown in blue where the dots and error bars indicate the mean results and $2\sigma$ and the dashed line is the fitted $m_\rho-m_\pi$ dependence \eqref{rhomassfit}. The green curve is the holographic results from \cite{Erdmenger:2020flu} and the orange curve is the fitted lattice results based on \cite{HadronSpectrum:2008xlg,Briceno:2016mjc,ExtendedTwistedMass:2019omo}. We plot these on top of the figure extracted from \cite{Pelaez:2010je,Pelaez:2010fj} using unitarized $\chi$PT (IAM) where results are compared with other lattice computations. See \cite{Pelaez:2010je,Pelaez:2010fj} for more references.}
	\label{latticerho}
\end{figure}

Restoring the physical unit, in fig. \ref{latticerho}, we show the GTB results of the dependence of $\rho$ mass on the pion mass. For comparison, in the same plot, we also show the results of holographic calculations from \cite{Erdmenger:2020flu} and the fitted lattice QCD data (in that reference) from \cite{HadronSpectrum:2008xlg,Briceno:2016mjc,ExtendedTwistedMass:2019omo}. In addition, we overlap the plots with the one from \cite{Pelaez:2010je} where several lattice results are compared with the unitarized $\chi$PT analysis. The $m_\rho-m_\pi$ behavior from GTB is in good agreement with the lattice results. The actual value of the mass differs because the choice of asymptotic form factor coefficients leads to a base value (the value for physical pion mass) that is higher than the actual $\rho$ mass. The plot also suggests our method to estimate our errors are in the correct ball park as they incorporate the gap to the lattice values.  In any case we find a fit:\footnote{The notation $833(06)$ here means $833\pm6$, and similarly for the results below.}
\beq\label{rhomassfit}
m_\rho \simeq 833 (06)\,\MeV + 0.00062 (06)\,\mpi^2
\eeq
which leads to $\mpi\simeq 492\,\MeV$ for the value when $\mpi=\half m_\rho$.

\begin{figure}[h!]
	\centering
	\includegraphics[width=0.75\textwidth]{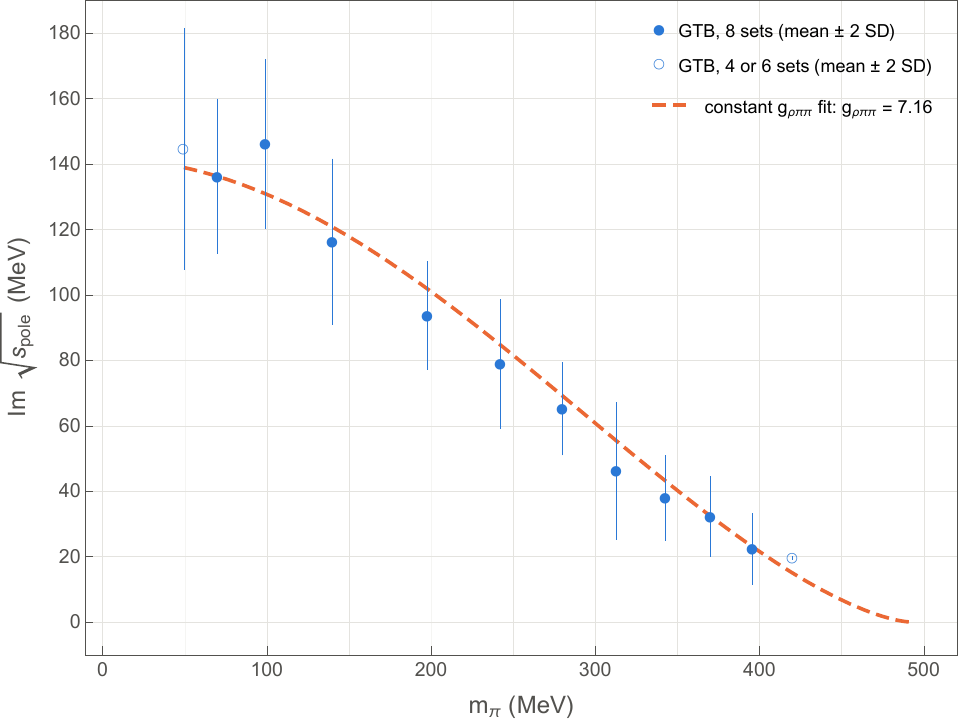}
	\caption{Consistency check of the $\text{Im}\sqrt{s_\text{pole}}$ with the half-width formula \eqref{phasespacegamma}. The blue points are the GTB numerical results of $\text{Im}\sqrt{s_\text{pole}}$ with $2\sigma$ error and the orange line is the computation using the formula \eqref{phasespacegamma} with the fitted $\rho$ mass \eqref{rhomassfit}. 
    }
	\label{gfit}
\end{figure}

We can perform another interesting cross check on our results. 
One expects the formula for the $\rho$ half-width  
\beq\label{phasespacegamma}
\mathrm{Im}\sqrt{s_{\rm pole}} = \frac{\Gamma_\rho}{2}
 = g_{\rho\pi\pi}^2\,\frac{m_\rho}{96\pi}\left(1-\frac{4m_\pi^2}{m_\rho^2}\right)^{3/2}
\eeq
where, in first approximation, the $\rho-\pi\pi$coupling can be considered independent of $\mpi$. Using \eqref{phasespacegamma} and the fit \eqref{rhomassfit}, we can fit the value of $g^2_{\rho\pi\pi}$ and then compute the imaginary part of the pole.  In figure \ref{gfit}, we show in orange the result of this computation with our best fit value $g^{\rho\pi\pi}=7.1(2)$. Our pole fitting data is the blue dots in figure \ref{gfit} which agree quite well with the recomputation of the orange curve.
For comparison putting the physical meson masses and width ($m_\rho=775$ MeV and $\Gamma_\rho=145$ MeV at $m_\pi=140$ MeV) into  \eqref{phasespacegamma} gives $g^{\rho\pi\pi}=5.96$.  \newpage

\subsection{The $\sigma$ meson, $S0$ channel}

\begin{figure}[h]
	\centering
	\includegraphics[width=\textwidth]{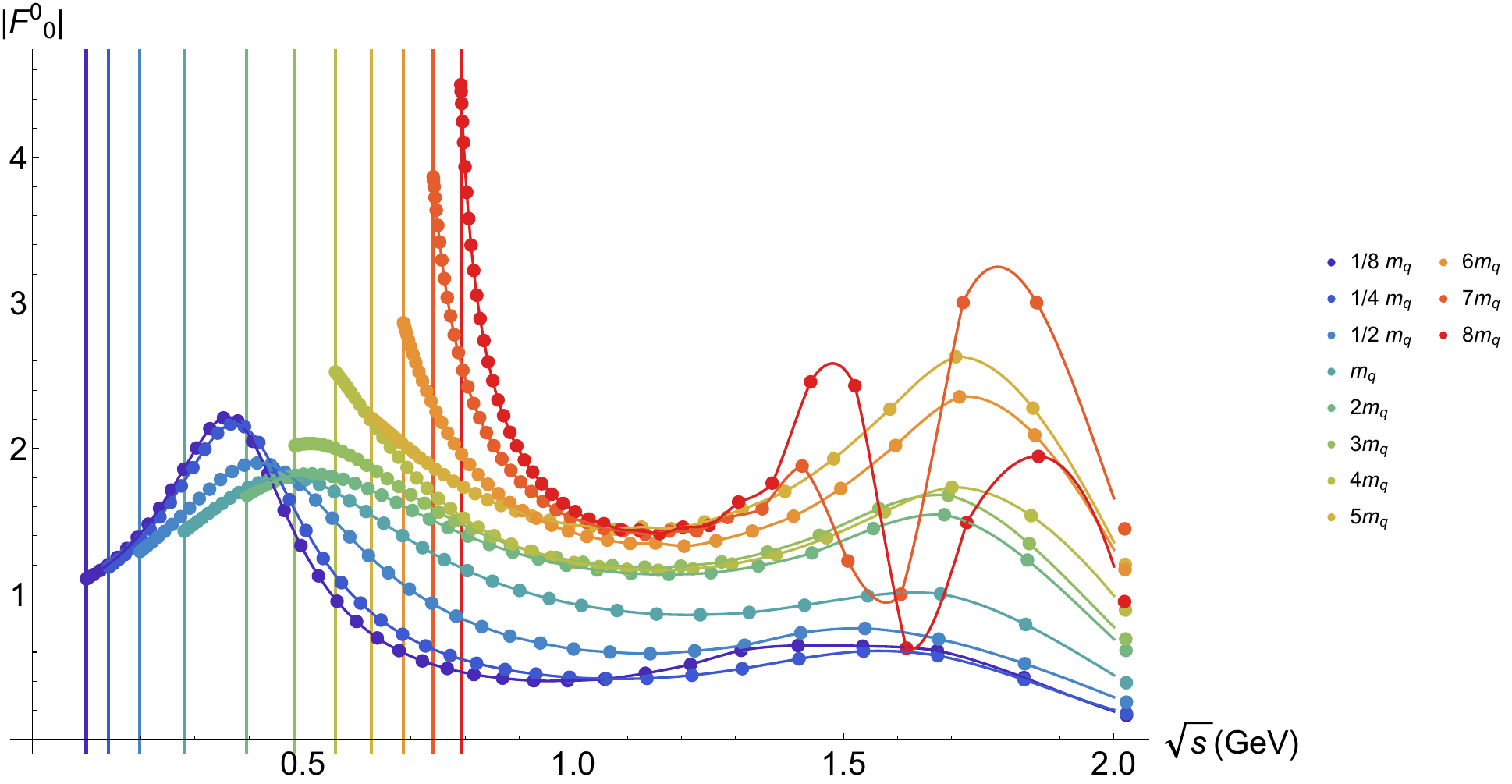}
	\caption{Modulus of scalar form factor $F^0_0$ as a function of energy $\sqrt{s}$ for varying quark mass. The gridlines with corresponding color indicates the moving threshold $\sqrt{s}=2m_{\pi}$. The legend corresponds to the range of $m_q$ values \eqref{mqlist} where we neglect the superscript $^{\text{phys}}$.}
	\label{FFS0}
\end{figure}

The $S0$ channel corresponds to isospin $I=0$ and angular momentum $\ell=0$. At the physical pion mass, the form factor has a clear, broad resonance that should be identified as the $\sigma$-meson. See fig. \ref{FFS0}. For larger pion mass, the $\sigma$ bump disappears (moving below threshold) and instead another resonance becomes more prominent at higher energy (which we will not study).

\begin{figure}[h!]
	\centering
	\includegraphics[width=\textwidth]{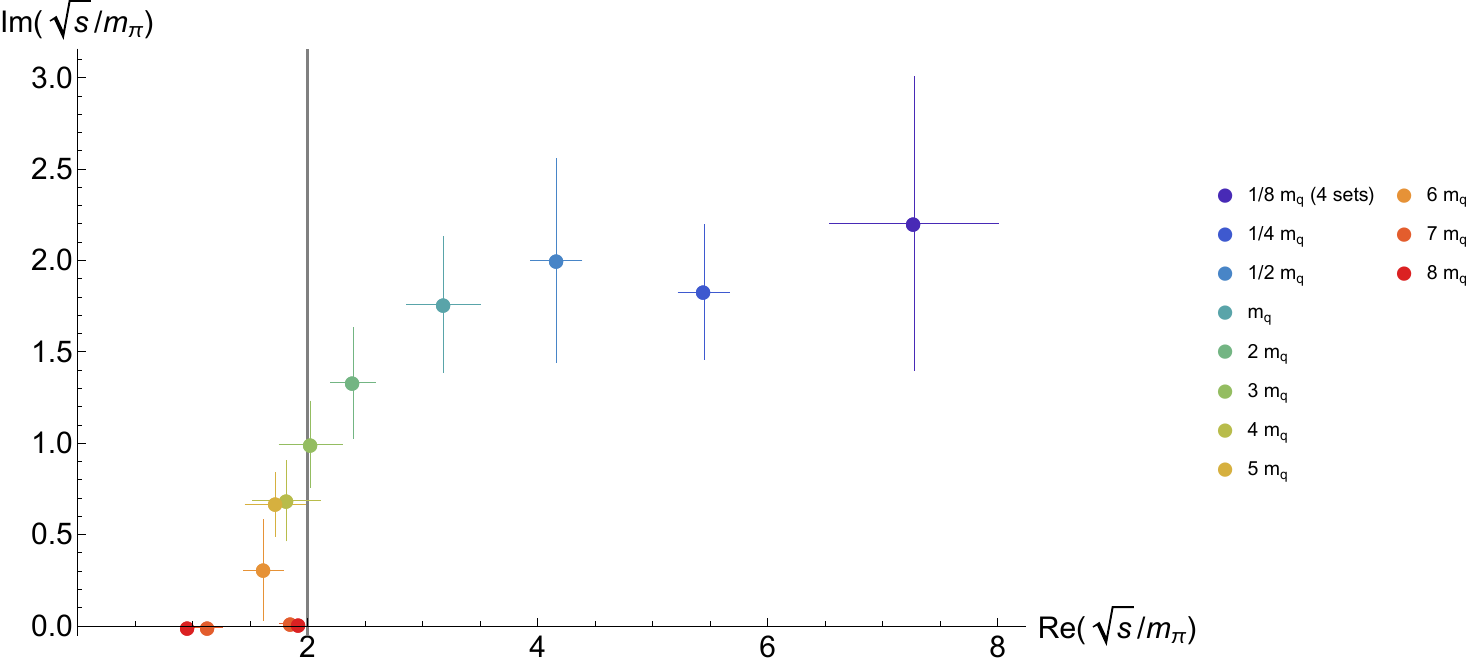}
	\caption{The location of the $\sigma$ meson pole in the complex plane of $\sqrt{s}$ for varying quark mass where $\sqrt{s_{\text{pole}}}=m_{\sigma}+i\Gamma_{\sigma}/2$. The plot is done in units of the pion mass where $m_\pi=1$. The results and $2\sigma$ error bars for each quark mass are obtained by averaging over the set \eqref{chival}. The legend corresponds to the range of $m_q$ values \eqref{mqlist} where we neglect the superscript $^{\text{phys}}$.}
	\label{sigmapole}
\end{figure}

Similar to before, the bump in the form factor can be understood as a pole of the form factor, and consequently of the amplitude on the second sheet of complex $s$. Unlike the sharp $\rho$ peak, the broad and disappearing $\sigma$ bump means the pole is far from the real axis. In the case of physical pion mass, the $\sigma$ pole was previously extracted in GTB in \cite{Cordoba:2025wnv}. Here, we extract the zero of $S^ 0_0(s)$ in the same way as described in the previous section to obtain the results for the set of quark mass \eqref{mqlist} (equivalently pion mass \eqref{mpilist}). The results of the location of the zero on the complex plane of $\sqrt{s}$ is shown in fig. \ref{sigmapole} where everything is written in the pion mass unit $m_\pi=1$. As the quark (or pion) mass increases, the pole moves towards threshold with a finite imaginary part. (There is another zero in the lower half plane due to real analyticity which we are not plotting.) The pair of poles then approach the real axis below the threshold and become a pair of virtual states. At higher quark mass $m_q=8m^{\text{phys}}_q$, one of the virtual states approaches the threshold from below.

\begin{figure}[h!]
	\centering
	\includegraphics[width=0.9\textwidth]{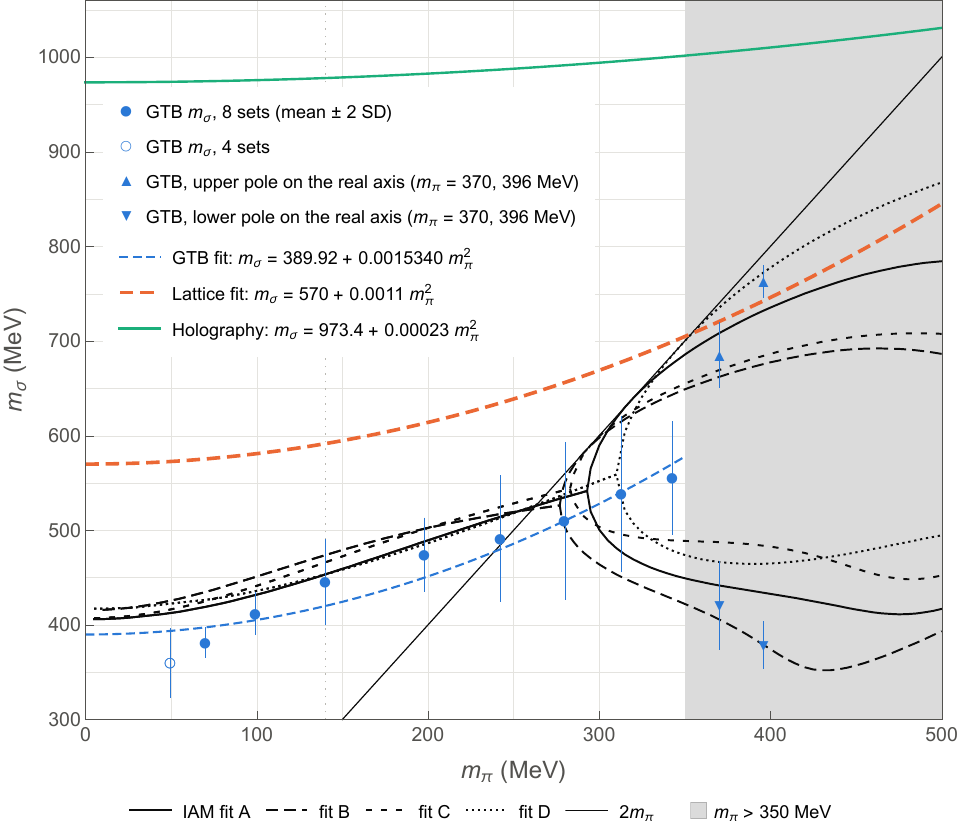}
	\caption{Comparison of the GTB results with lattice, holographic and unitarized $\chi$PT results. The GTB computations are shown in blue where the dots and error bars indicate the mean results and $2\sigma$ and the dashed line is the fitted $m_\sigma-m_\pi$ dependence \eqref{rhomassfit}. The green curve is the holographic results from \cite{Erdmenger:2020flu} and the orange curve is the fitted lattice results based on \cite{HadronSpectrum:2008xlg,Briceno:2016mjc,ExtendedTwistedMass:2019omo}. We plot these on top of the figure extracted from \cite{Pelaez:2010je} using unitarized $\chi$PT (IAM) where results are compared with other lattice computations. See \cite{Pelaez:2010je} for more references.}
	\label{sigmapimass}
\end{figure}

Using \eqref{polemG}, we can extract the mass of the $\sigma$ meson from the real part of the pole location. Restoring the physical unit, in fig. \ref{sigmapimass}, we plot the $\sigma$ mass against the pion mass in physical units. We include the comparison with the holographic results from \cite{Erdmenger:2020flu} as well as the fitted lattice QCD data from \cite{HadronSpectrum:2008xlg,Briceno:2016mjc,ExtendedTwistedMass:2019omo}.
The fitted result gives
\begin{equation}
    m_\sigma=390 (11)\text{ MeV}+0.0015(02)m^2_\pi
\end{equation}
Our results generate a lighter $\sigma$ than the lattice but with very similar $m_\sigma - m_\pi$ dependence. 

\newpage

\begin{figure}[h]
	\centering
	\includegraphics[width=12cm]{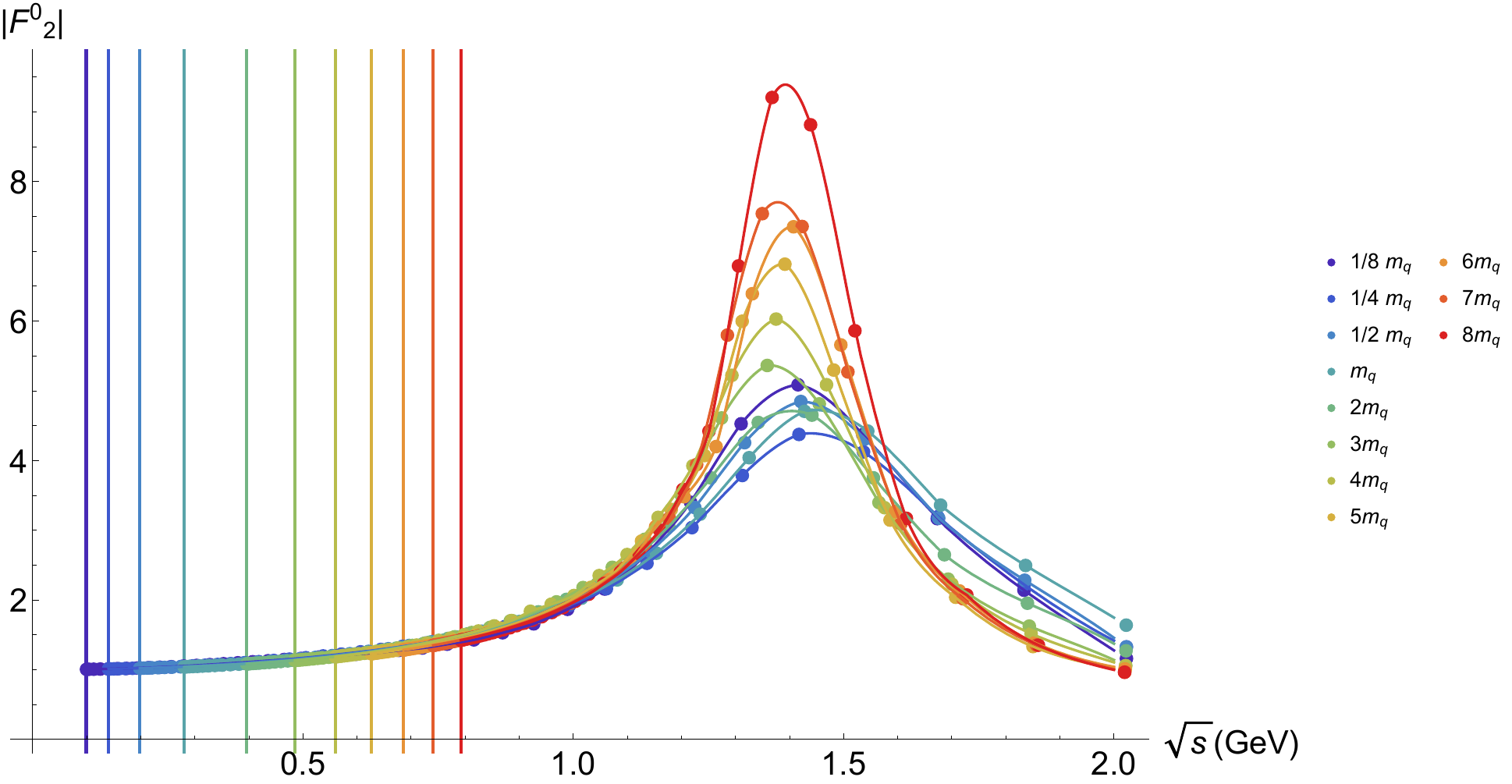}
	\caption{Modulus of tensor form factor $F^0_2$ as a function of energy $\sqrt{s}$ for varying quark mass. The gridlines with corresponding color indicates the moving threshold $\sqrt{s}=2m_{\pi}$. The legend corresponds to the range of $m_q$ values \eqref{mqlist} where we neglect the superscript $^{\text{phys}}$.}
	\label{FFD0}
\end{figure}

\begin{figure}[h!]
	\centering
	\includegraphics[width=12cm]{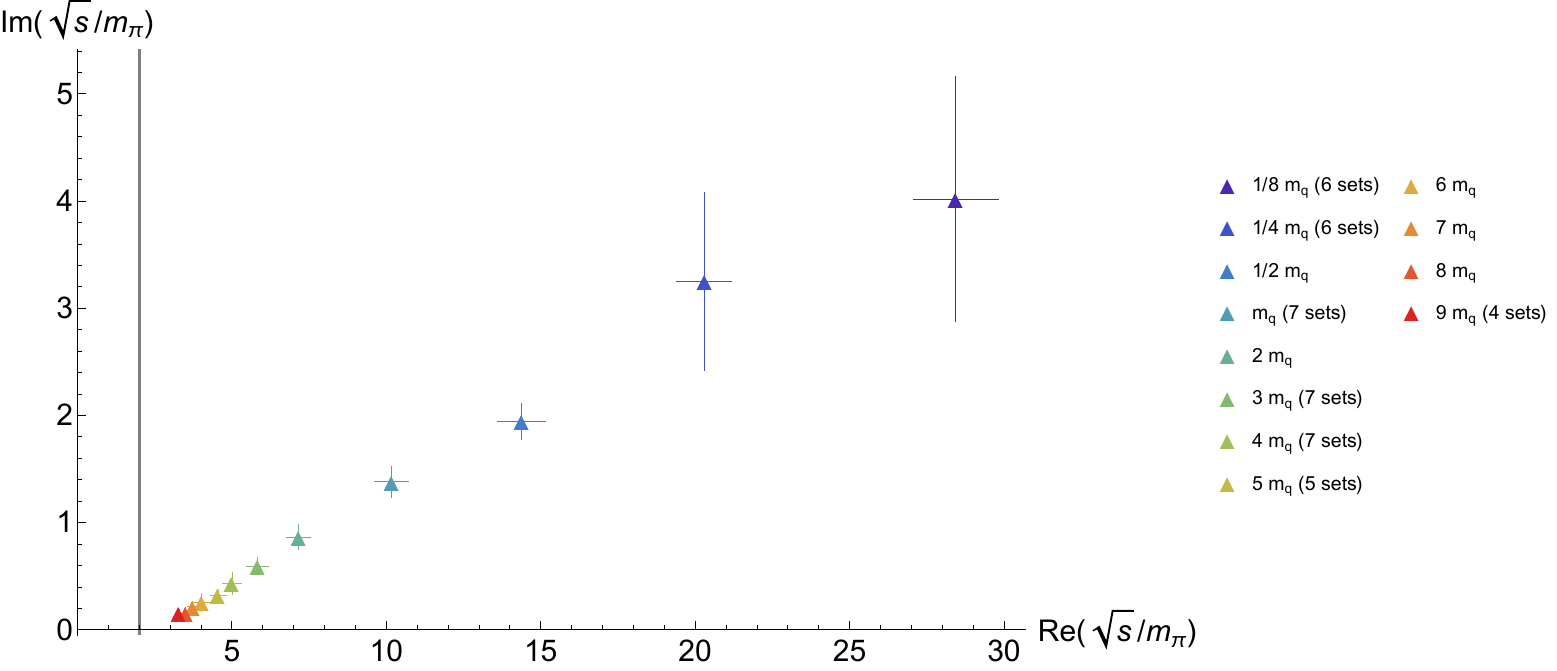}
	\caption{The location of the $f_2$ meson pole in the complex plane of $\sqrt{s}$ for varying quark mass where $\sqrt{s_{\text{pole}}}=m_{f_2}+i\Gamma_{f_2}/2$. The plot is done in units of the pion mass where $m_\pi=1$. The results and $2\sigma$ error bars for each quark mass are obtained by averaging over the set \eqref{chival} where for some quark mass values, due to limited numerical precision, we extract the zero from a smaller subset as indicated. The legend corresponds to the range of $m_q$ values \eqref{mqlist} where we neglect the superscript $^{\text{phys}}$.}
	\label{f2pole}
\end{figure}

\subsection{The $f_2$ meson, $D0$ channel}

Finally, we consider the $D0$ channel corresponds to isospin $I=0$ and angular momentum $\ell=2$. The form factor has a clear $f_2$ resonance for physical pion mass that remains for all other values of the pion mass (that we considered). See fig. \ref{FFD0}. With increasing pion mass, the $f_2$ mass in the physical unit remains stable, and the peak gets narrower. This is similar to the case of the $\rho$ resonance where as the mass of the pion increases there is less phase space available for the decay so the resonance becomes narrower. Similar to before, we can extract the second-sheet poles in the partial amplitudes $S^0_2(s)$ and we plot the averaged position and errors of the pole in fig. \ref{f2pole} (in units of the pion mass).

\begin{figure}[h!]
	\centering
	\includegraphics[width=0.8\textwidth]{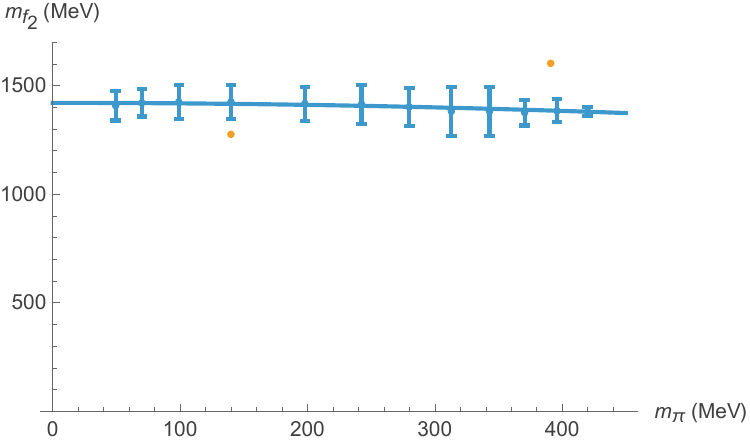}
	\caption{The  $f_2$ meson mass plotted against the pion mass in physical units (MeV). the best fit line $m_{f_2} = 1420 -0.00023 m_\pi^2$ is also shown. The orange points are the physical point $m_\pi=140\text{MeV}$, $m_{f_2}=1275$MeV and the value in  \cite{Briceno:2017qmb} -- $m_\pi=1602(10)$MeV at $m_\pi = 391$MeV. }
	\label{f2mass}
\end{figure}

\begin{figure}[h!]
	\centering
	\includegraphics[width=0.8\textwidth]{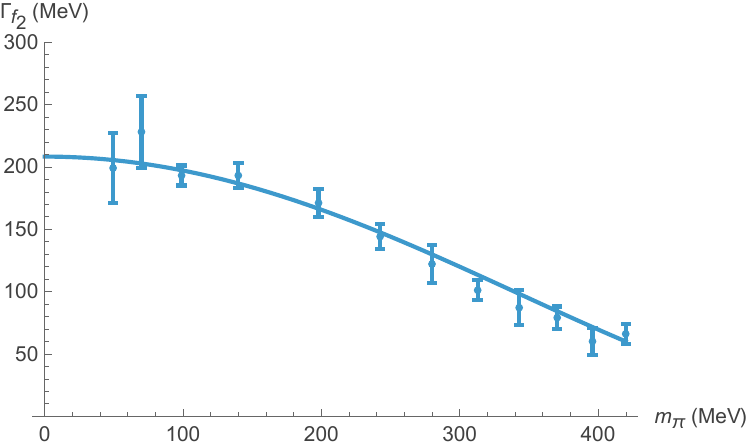}
	\caption{The  $f_2$ meson width plotted against the pion mass in physical units (MeV). The best fit line of the form in eq. (\ref{gf2}) with the fit for $m_{f_2}$ in eq. (\ref{f2fit}) and the best fit value $g_{f_2 \pi \pi} = 14.8$ is also shown. }
	\label{f2gamma}
\end{figure}

In Fig. \ref{f2mass} we display the $f_2$ mass as a function of $m_\pi$ in physical units and also display the known physical point and another point taken from the lattice. The best fit to our data is
\begin{equation} \label{f2fit}
    m_{f_2}=1420(04)\text{ MeV}-0.00023(03) m^2_\pi
\end{equation}
Clearly this does not capture the expected rise in mass of the state with rising quark mass. The output value does though lie between the range given by the physical point and a higher quark mass point from the lattice \cite{Briceno:2017qmb} (shown by orange points in fig. \ref{f2mass}). Higher mass resonances are likely harder to predict using the GTB and this is an example where there is some useful information but not the precision to capture the quark mass dependence.

Now, for the $f_2$ width we can use  the formula taken from \cite{EWERZ201431}\footnote{The $f_2$ pion coupling is a dimension five operator of the schematic form $\cL \sim \frac{g_{f_2\pi\pi}}{M_0} \phi \partial \pi\partial \pi$. In \cite {EWERZ201431} a reference mass $M_0=1\; \GeV$ is introduced to define a dimensionless coupling.}
\beq \label{gf2}
\Gamma(f_2\rightarrow\pi^0\pi^0) = \half\Gamma(f_2\rightarrow \pi^+\pi^-) = \frac{m_{f_2}}{960\pi} |g_{f_2\pi\pi}|^2 \left(\frac{m_{f_2}}{M_0}\right)^2 \left(1-\frac{4m_\pi^2}{m_{f_2}^2}\right)^\frac{5}{2}
\eeq
In Fig. \ref{f2gamma} we plot our data for the width against the pion mass in physical units. We also use eq. (\ref{gf2}) with the fit eq. (\ref{f2fit}) to fit $g_{f_2\pi\pi}$ for which we find the value 14.8(2). This self consistency check also works well for our data. (For comparison if one uses the physical data $m_\pi=140$ MeV, $m_{f_2}=1,270$ MeV, $\Gamma_{f_2}=150$ MeV in eq. (\ref{gf2}) then one finds the value $g_{f_2 \pi \pi}=15.8$.) \newpage

\subsection{Summary of spectra \label{sum}}

Our extracted data for the $\rho$ meson is\\

\begin{tabular}{c|c|c|cc}
$m_q/m_q^{\rm phys}$ & $m_\pi$ (MeV) & data sets & $m_\rho$ (MeV)  & $\Gamma_\rho$ (MeV) \\ \hline
&&&&\\
1/8& 49.5 & 6&$807 \pm 113$&$145 \pm 37$\\
1/4&70&8&$827 \pm 102$&$136 \pm 24$\\
1/2&99&8&$861 \pm 113$&$146 \pm 26$\\
1&140&8&$861 \pm 76$&$116 \pm 25$\\
2&198&8&$864 \pm 58$&$94 \pm 16$\\
3&242.5&8&$874 \pm 70$&$79 \pm 20$\\
4&280&8&$880 \pm 79$&$65 \pm 14$\\
5&313&8&$893 \pm 71$&$46 \pm 21$\\
6&343&8&$904 \pm 51$&$38 \pm 13$\\
7&370.5&8&$914 \pm 30$&$32 \pm 12$\\
8&396&8&$927 \pm 31$&$22 \pm 10$\\
9&420&4&$948 \pm 0.8$&$20 \pm 0.5$
\end{tabular}\\

\noindent Again the best fit for the mass  is
\beq 
m_\rho \simeq 833 (06)\,\MeV + 0.00062 (04)\,\mpi^2
\eeq
From the width data we determined $g_{\rho \pi \pi} = 7.1(2)$.\\

The data for the $\sigma$ meson is\\

\begin{tabular}{c|c|c|cc}
$m_q/m_q^{\rm phys}$ & $m_\pi$ (MeV) & data sets & $m_\sigma$ (MeV)  & $\Gamma_\sigma$ (MeV) \\ \hline
1/8& 49.5 & 4&$360 \pm 36$& $109 \pm 20$\\
1/4&70&8&$381 \pm 16$& $ 128 \pm 13$\\
1/2&99&8&$412 \pm 22$&$199\pm 28$\\
1&140&8&$445 \pm 45$& $246 \pm 26$\\
2&198&8&$474 \pm 39$&$263 \pm 30$\\
3&242.5&8&$491 \pm 67$&$241 \pm 29$\\
4&280&8&$509 \pm 83$&$192 \pm 31$\\
5&313&8&$538 \pm 83$&$209 \pm 27$\\
6&343&8&$555 \pm 60$&$105 \pm 47$\\
7&370.5&8&$420 \pm 46$&$-4 \pm 1$\\
&&&685 $\pm$ 34&$8 \pm 1$\\
8&396&8&$379 \pm 25$&$-3 \pm 1.5$\\
&&&763 $\pm$ 17&$5 \pm 3$
\end{tabular}\\
The best fit here is
\begin{equation}
    m_\sigma=390(11)\text{ MeV}+0.0015(02)m^2_\pi
\end{equation}

The data for the $f_2$ meson is\\

\begin{tabular}{c|c|c|cc}
$m_q/m_q^{\rm phys}$ & $m_\pi$ (MeV) & data sets & $m_{f_2}$ (MeV)  & $\Gamma_{f_2}$ (MeV)  \\ \hline
1/8& 49.5 & 6&$1407 \pm 68$&$199 \pm 28$\\
1/4&70&6&$1420 \pm 63$& $228 \pm 29$\\
1/2&99&8&$1424 \pm 77$&$193 \pm 8$\\
1&140&7&$1424 \pm 77$& $193 \pm 10$\\
2&198&8&$1416 \pm 78$&$171 \pm 11$\\
3&242.5&7&$1414 \pm 90$&$144 \pm 10$\\
4&280&7&$1401 \pm 87$&$122 \pm 15$\\
5&313&5&$1426 \pm 82$&$101 \pm 8$\\
6&343&8&$1381 \pm 113$&$87 \pm 14$\\
7&370.5&8&$1375 \pm 59$&$79 \pm 9$\\
8&396&8&$1383 \pm 53$&$60 \pm 11$\\
9&420&4&$1380\pm 21$& $66 \pm 8$
\end{tabular}\\

The best fit here is
\begin{equation}  
    m_{f_2}=1420(04)\text{ MeV}-0.00023(03) m^2_\pi
\end{equation}
From the width data we determined that $g_{f_2 \pi \pi} = 14.8 (2)$

\section{Low energy parameters}\label{lec}

In the low energy region, the theory is described by the chiral effective field theory and the low energy dynamics are captured by a set of low energy parameters: low energy constants, pion radii, scattering lengths and effective range parameters. In this section we extract these quantities from the GTB results and show their $m_\pi$ dependence.

\subsection{Low energy amplitude and pion radii}

The first low energy parameter is the quartic coupling of pions defined nonperturbatively as
\begin{equation} \label{def_lambda}
    \lambda=\frac{1}{32\pi}\mathcal{M}(\pi^0\pi^0\to\pi^0\pi^0)|_{s=t=u=\frac{4}{3}m_\pi^2}=\frac{3\pi}{4}A\bigg(\frac{4}{3}m_\pi^2,\frac{4}{3}m_\pi^2,\frac{4}{3}m_\pi^2\bigg)
\end{equation}
Its generic bound from analyticity, crossing and unitarity have been studied extensively in the modern S-matrix bootstrap approach \cite{Paulos:2017fhb,LM,He:2021eqn,Guerrieri:2021tak,Chen:2022nym,EliasMiro:2022xaa}. Moreover, the 1-loop chiral Lagrangian is characterized by a set of coefficients $\ell_j$ that allows to introduce the renormalized low energy constants $\bar{l}_j$ defined at a renormalization scale $\mu=m_\pi$ \cite{GASSER1984142}. At very low energy, the amplitude can be expanded in $\lambda$ to 1-loop as \cite{He:2025gws}
\beq\label{Afit}
A(s,t,u) = \frac{4}{\pi}\lambda (s-1) + \frac{4\lambda^2}{3\pi^2}\left[A_1(s,t,u)+A_2(s,t,u)-3(s-1)(\bar{A}_1+\bar{A}_2-3\bar{\ell}_3)-3\bar{\ell}_3\right] 
\eeq
 where
 \begin{subequations}
     \beqa
 \!\!\!\!\!\!\!\!A_1(s,t,u) =&&\!\!\!\!\!\!\!\! 3(s^2-m_{\pi}^4)I_f(s)   + (t(t - u) - 2tm_{\pi}^2 + 4um_{\pi}^2 - 2m_{\pi}^4)I_f(t)  \\ 
 && + (u(u - t) - 2um_{\pi}^2 + 4tm_{\pi}^2 - 2m_{\pi}^4)I_f(u) \non \\
 \!\!\!\!\!\!\!\!A_2(s,t,u) =&&\!\!\!\!\!\!\!\! 2\left(\bar{\ell}_1 - \frac{4}{3}\right)(s - 2m_{\pi}^2)^2 + \left(\bar{\ell}_2 - \frac{5}{6}\right)(s^2 + (t - u)^2) - 12 sm_{\pi}^2 + 15m_{\pi}^4 \nonumber
 \eeqa
 \end{subequations}
 with
 \beq
 I_f(s) = \beta(s)\ln\left(\frac{\beta(s) - 1}{\beta(s) + 1}\right)+ 2,\;\;\;\;
 \beta(s) = \sqrt{1-\frac{4m_{\pi}^2}{s}}
 \eeq
and $\bar{A}_{1,2}=A_{1,2}(\frac{4}{3},\frac{4}{3},\frac{4}{3})$.

From the GTB results, we then evaluate the amplitudes in a set of points in the very low energy region and fit the results to the form of \eqref{Afit} to extract the low energy parameters $\lambda,\bar{l}_{1,2,3}$ for each value of $m_\pi$.

In addition, the low energy constants $\bar{l}_{4,6}$ are encoded in the low energy expansion of the scalar and vector form factors as
\begin{subequations}
\begin{eqnarray}
        F^0_0(s) &=& F^0_0(0)\left(1+ {2 \lambda \over \pi}
 s \left(\bar{\ell_4} - {13
\over 12} \right)  + O(\lambda^2)\right)\\
        F^1_1(s) &=& 1+ { \lambda \over 3 \pi}
 s \left(\bar{\ell_6} -1\right)  + O(\lambda^2)
\end{eqnarray}
\end{subequations}
and can be extracted from the GTB results.

\begin{figure}[h!]
\centering
	\includegraphics[width=0.9\textwidth]{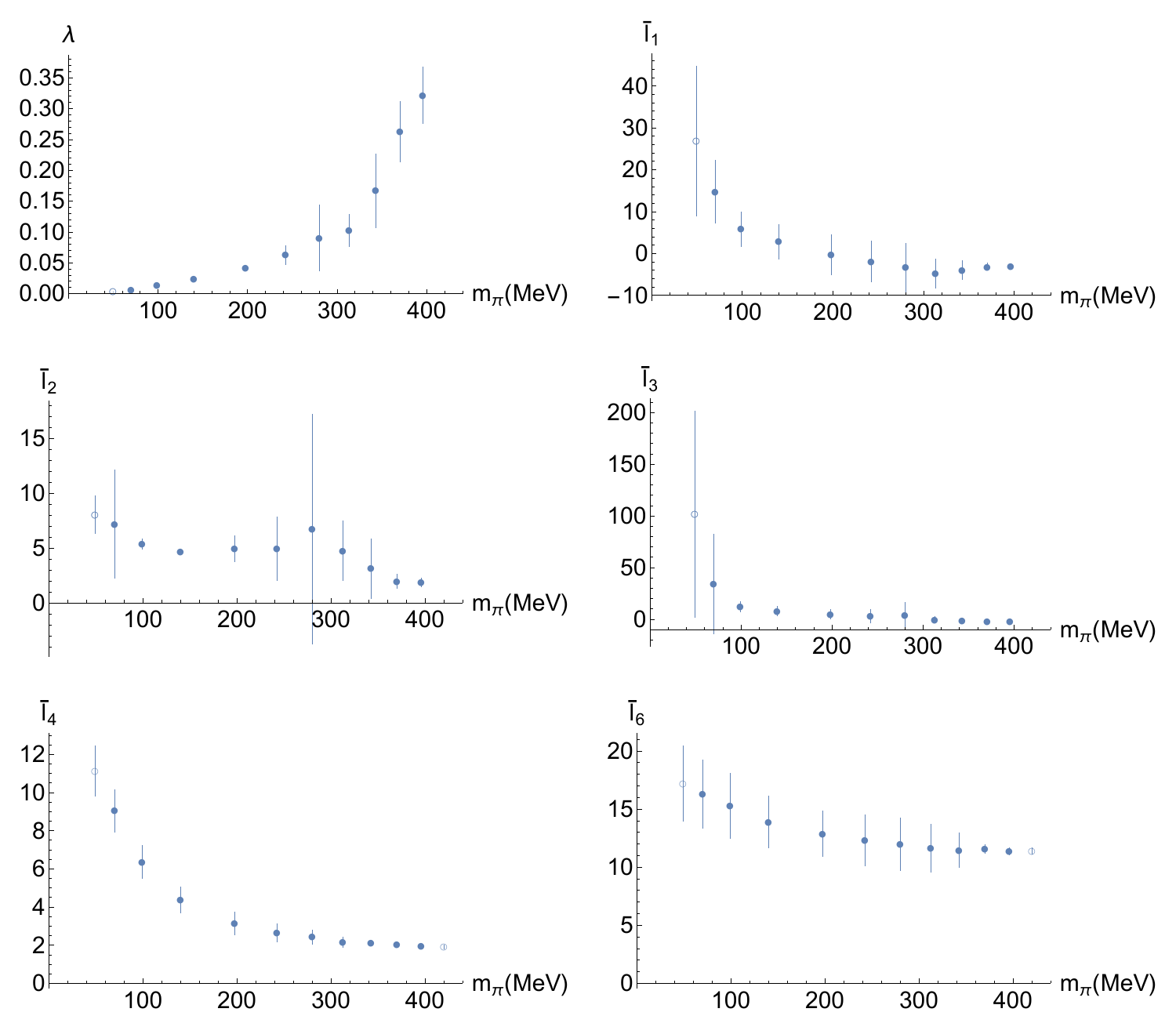}
	\caption{The low energy constants $\lambda,\bar{l}_{1,2,3,4,6}$ for varying pion mass $m_\pi$. The dots and error bars indicate the mean and $2\sigma$ over the set \eqref{chival}.}
	\label{LEC}
\end{figure}

In fig. \ref{LEC}, we show the results of the low energy constants $\lambda,\bar{l}_{1,2,3,4,6}$ as function of the pion mass $m_\pi$.

\begin{figure}[h!]
\centering
	\includegraphics[width=\textwidth]{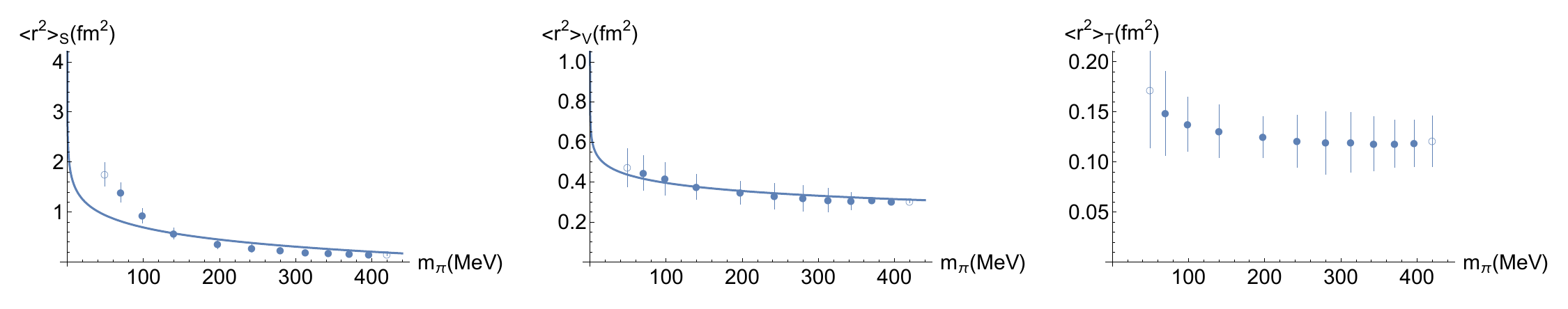}
	\caption{The scalar, vector and tensor pion radii $\langle r^2\rangle_{S,T,U}$ in fm$^2$ for varying pion mass $m_\pi$. The dots and error bars indicate the mean and $2\sigma$ over the set \eqref{chival}. The continuous curves are the expected expression \eqref{rchiPT} from leading order $\chi$PT, where there is a large deviation for the scalar form factor.}
	\label{radii}
\end{figure}

Note that the low energy expansion of the form factors also give the pion charge radii:
\begin{subequations}
\begin{eqnarray}
        F_0^0(s) &=& F_0^0(0) \left( 1 + {1 \over 6} ~ s ~ \langle r^2\rangle _\pi^S+ ...\right)\\
    F_1^1(s) &=& 1 + {1 \over 6} ~ s ~ \langle r^2\rangle _\pi^V+ ...\\
    F_0^2(s) &=& 1 + {1 \over 6} ~ s ~ \langle r^2\rangle _\pi^T+ ...
\end{eqnarray}
\end{subequations}
The pion radius for the scalar and vector currents should decrease as the pion mass increases - that is, the pion cloud should become larger as the pion becomes lighter. From $\chi$PT\, we expect, at lowest order, for small pion mass \cite{GASSER1984142},
\beqa\label{rchiPT}
\langle r^2\rangle _\pi^S(\mpi) &=& \langle r^2\rangle _\pi^S(\bmpi)-\frac{3}{8\pi^2 \fpi^2} \ln\frac{\mpi^2}{\bmpi^2} \\
\langle r^2\rangle _\pi^V(\mpi) &=& \langle r^2\rangle _\pi^V(\bmpi)  -\frac{1}{16\pi^2 \fpi^2} \ln\frac{\mpi^2}{\bmpi^2}
\eeqa
where $\bar{m}_\pi$ stands for a different value of $m_\pi$.
In fig. \ref{radii}, we show the pion radii as function of $m_\pi$ and compare with the expression \eqref{rchiPT}. While the vector form factor agrees relatively well with \eqref{rchiPT}, the scalar form factor has a larger deviation from the leading $\chi$PT results.

\subsection{Scattering lengths and effective range}

Near threshold, the partial waves can be characterized by the scattering lengths and the effective range parameters defined from an expansion of the real part of the partial waves:
\begin{equation}\label{scatlen}
 \text{Re}f^I_{\ell}(s)\overset{k\to 0}{\simeq}\frac{2}{\pi}\left(\frac{k^{2}}{m_\pi^2}\right)^{\ell}\big(a^I_{\ell}+b^I_{\ell}\frac{k^2}{m_\pi^2}+\ldots\big),\;\; k=\frac{\sqrt{s-4m_{\pi}^2}}{2}
\end{equation}
where we have used pion mass as unit.
At tree-level, the Weinberg model \cite{PhysRevLett.17.616} already gives reasonable values for the $S0$, $S2$ and $P1$ scattering lengths (see  also \eg\ \cite{donoghue_golowich_holstein_2014}).

\begin{equation}
\begin{aligned}
        &a_0^0 = \frac{7m^2_\pi}{32 \pi f_\pi^2}, ~~~ a_0^2= -  \frac{m^2_\pi}{16\pi f_\pi^2},\\
        &a_1^1 = \frac{m^2_\pi}{24 \pi f_\pi^2},~~~~ b_0^0 = \frac{m^2_\pi}{4\pi f_\pi^2},~~~ b_0^2 = - \frac{m^2_\pi}{8\pi f_\pi^2},\\ 
        &a_2^0 = a_2^2= b^1_1=0.\label{Weinbergab}
\end{aligned}
\end{equation}
These are our basic expectation and are all linear in the quark mass ($m_\pi^2 \propto m_q$). Notice that, for $a^1_1$, $b^0_0$, $b^2_0$ this is  due to the definition \eqref{scatlen} that uses pion mass units, otherwise they would be independent of $m_\pi$. Note we are assuming $f_\pi$ is independent of $m_q$ in our analysis. 

\begin{figure}[h!]
\centering
	\includegraphics[width=0.86\textwidth]{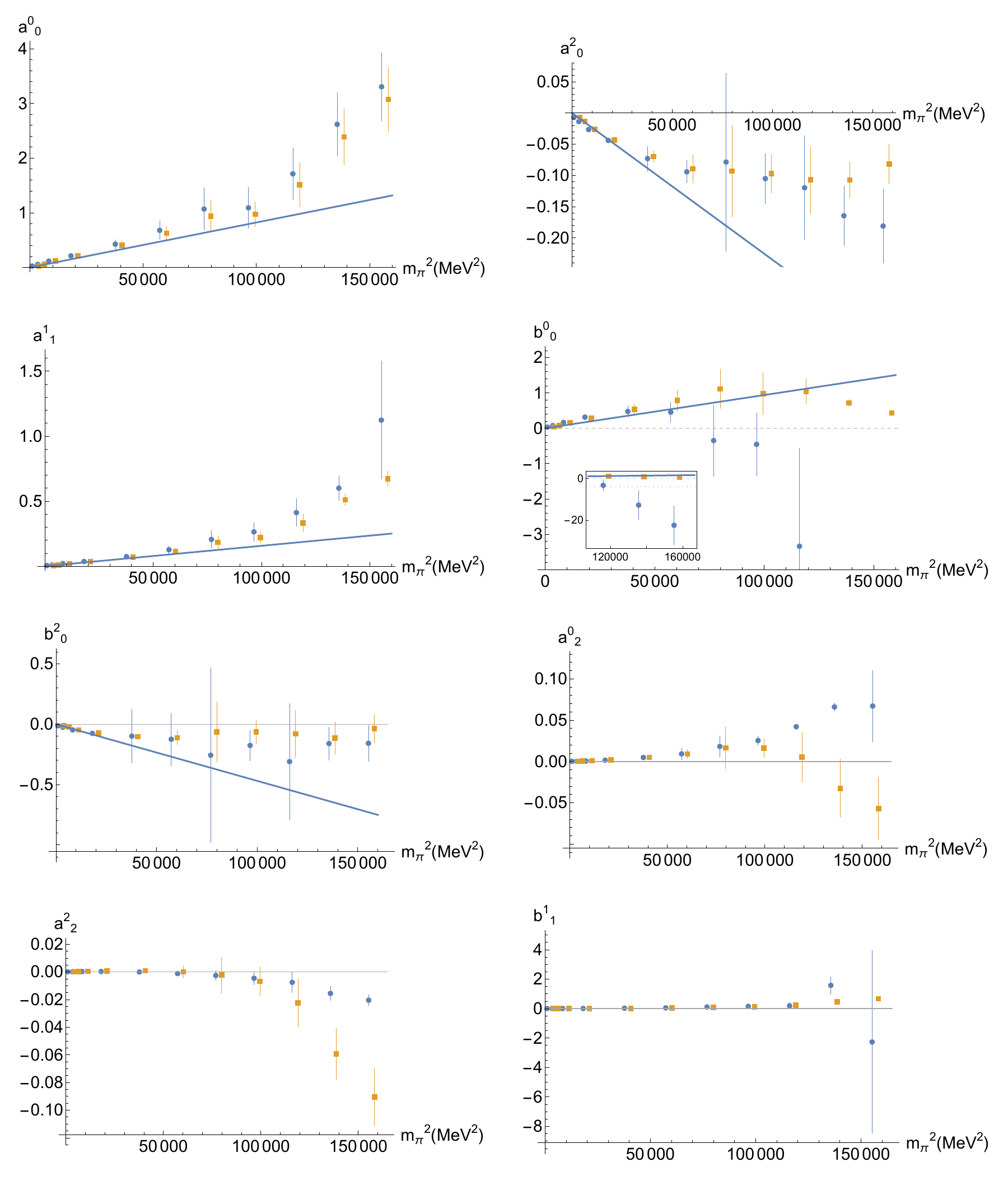}
	\caption{The mean and $2\sigma$ error of the scattering lengths $a^I_\ell$ and effective range parameters $b^I_\ell$ from direct near-threshold fit \eqref{scatlen} (blue dots) and 1-loop recomputation using the low energy constants \eqref{larelation} (orange dots). Their approximate agreement (in particular at smaller values of $m_\pi$) is an indication of the consistency of the computation. The blue lines show the leading $\chi$PT expectation from Weinberg model.}
	\label{abplot}
\end{figure}

From GTB results, we can evaluate the three $f_\ell^I$ at a list of points near threshold and fit them into the form \eqref{scatlen} to extract the $a^I_\ell,b^I_\ell$ and plot them for varying $m_\pi^2$. 
We display these fit as blue points in fig. \ref{abplot}. For comparison, the blue lines are the Weinberg model prediction \eqref{Weinbergab}. In all cases there is a clear departure from the chiral Lagrangian precition at larger $m_\pi$. This indicates that higher order corrections in $\lambda$ are important.

At 1-loop the results in chiral perturbation theory are corrected. The scattering lengths and effective range parameters depend on the low energy constants (LEC) $\bar{l}_{1,2,3}$ as follows \cite{GASSER1984142,He:2025gws}:
\begin{eqnarray}
a^{(0)}_0&=&7 \lambda +\frac{\lambda ^2}{9 \pi } \left[64 \bar{\ell}_1+128 \bar{\ell}_2+144 \bar{\ell}_3+84+231 \sqrt{2} \tan ^{-1}(2
   \sqrt{2})\right]+\mathcal{O}(\lambda^3)\nonumber\\
   a^{(2)}_0&=&-2 \lambda +\frac{2\lambda ^2}{9 \pi } \left[32 \bar{\ell}_1+64 \bar{\ell}_2-36 \bar{\ell}_3+42-33 \sqrt{2} \tan ^{-1}(2 \sqrt{2})\right]+\mathcal{O}(\lambda^3)\nonumber\\
   a^{(1)}_1&=&\frac{4 \lambda }{3}+\frac{2 \lambda ^2}{27 \pi }\left[-64 \bar{\ell}_1+16 \bar{\ell}_2+54 \bar{\ell}_3-131+66 \sqrt{2} \tan ^{-1}(2
   \sqrt{2})\right]+\mathcal{O}(\lambda^3)\nonumber\\
   b^{(0)}_0&=&8 \lambda +\frac{4\lambda ^2}{9 \pi } \left[80 \bar{\ell}_1+112 \bar{\ell}_2+54 \bar{\ell}_3-105+66 \sqrt{2} \tan ^{-1}(2
   \sqrt{2})\right]+\mathcal{O}(\lambda^3)\nonumber\\
   b^{(2)}_0&=&-4 \lambda +\frac{2\lambda ^2}{9 \pi } \left[64 \bar{\ell}_1+176 \bar{\ell}_2-54 \bar{\ell}_3+51-66 \sqrt{2} \tan ^{-1}(2 \sqrt{2})\right]+\mathcal{O}(\lambda^3)\nonumber\\
   a_2^{(0)} &=& \frac{32\lambda^2}{45\pi}\left[\bar{\ell}_1+4\bar{\ell}_2-\frac{53}{8}\right]+\mathcal{O}(\lambda^3) \nonumber\\
a_2^{(2)} &=& \frac{32\lambda^2}{45\pi}\left[\bar{\ell}_1+\bar{\ell}_2-\frac{103}{40}\right]+\mathcal{O}(\lambda^3)\label{larelation}
\end{eqnarray}
We can then use the set of LEC $\lambda,\bar{l}_{1,2,3}$ extracted in the previous section and plotted in fig. \ref{LEC} to recompute the scattering lengths and effective range parameters for each $m_\pi$. The results are plotted in fig. \ref{abplot} in orange. The comparison of the direct fit (blue points) and 1-loop recomputation (orange points) gives an idea of the self-consistency of the GTB computation. As seen in fig. \ref{abplot}, the orange 1-loop recomputed results are mostly closer to the blue points than the tree level fit showing reasonable internal consistency of the results. The better agreement at smaller pion mass is reasonable since the 1-loop correction should work better at smaller pion mass.

\section{Conclusions}

The Gauge Theory Bootstrap approach provides a way to study the strongly coupled dynamics of QCD with controllable UV theory parameters. The method fixes $N_c,N_f,m_q$ and $\alpha_s$ in the UV and $m_\pi$ and $f_\pi$ in the IR. The Watsonian iteration then enforces saturation of unitarity on the $2 \rightarrow 2$ pion scattering amplitude, the $\pi \pi$ vacuum form factor and the current correlator. In this work we study the  approach for a range of quark mass or equivalently pion mass, from close to the chiral limit to the heavy pion mass limit where the resonances approach being stable. We extracted the $\sigma$, $\rho$ and $f_2$ resonance and studied how their masses and widths depend on the pion mass - these results are summarized in section \ref{sum}.  The results compare reasonably well with previous determinations from lattice or unitarized $\chi$PT, showing the strength of the method even away from the physical quark mass value. In addition, we also computed low energy quantities -- low energy constants, pion radii, scattering lengths and effective range -- and show their $m_\pi$ dependence. The results followed expectations based on physical grounds or on previous work. The consistency of all these results provide a further non--trivial check of the Gauge Theory Bootstrap approach. Recent work \cite{Cordoba:2026wsy} also examined the Gauge Theory Bootstrap approach for varying number of colors $N_c$ with consistent results. In the future, the method may allow us to move even further from QCD by changing the number of flavors $N_f$, giving an understanding of the space of confining strongly coupled gauge theories without the computational cost of lattice gauge theory methods.

\section*{Acknowledgements}

We are grateful to Kostas Rigatos for collaboration at the early stage of the project, and Max Hansen and Fernando Romero-Lopez for useful discussions. N.E.’s work was supported by the STFC consolidated grants
ST/X000583/1 and ST/C004240/1. 

\newpage

\appendix

\section{$m_q$-dependent numerical discretization}

The numerical GTB implementation we used here is similar to the one described in appendix C of recent work \cite{Cordoba:2026wsy}. One sutblety in the current $m_\pi$-dependent GTB computations is that, since we always take $m_\pi=1$ to set the energy unit, we are changing the energy unit for each GTB run for the set of quark masses \eqref{mqlist} or pion masses \eqref{mpilist}. To guarantee the numerical precision in the strongly coupled region, we then use a different discretization (in squared energy $s$) for each value of $m_\pi$. 
In this appendix, we describe this $m_q$-dependent discretization and refer interested readers to appendix C of \cite{Cordoba:2026wsy} for other detailed aspects of the GTB numerics. The numerical program for generating the results of this paper is available at the github repository \href{https://github.com/hyfysics/gauge-theory-bootstrap}{gauge-theory-bootstrap}.

\subsection{Discretized variables and constraints}

The amplitude and form factors are analytic functions on the cut-$\nu$ planes\footnote{Here $\nu$ stands for the Mandelstam invariants $s,t,u$.} with cuts $\nu>4m_\pi^2$. For each value of $m_\pi$, numerically we set $m_\pi=1$ to be the energy unit so we have the cuts starting from $4$. Such a cut plane can be mapped to the unit disk through the conformal map:
\beq
z(\nu) = \frac{\sqrt{4-\nu_0}-\sqrt{4-\nu}}{\sqrt{4-\nu_0}+\sqrt{4-\nu}}, \ \ \ \ z(x+i\epsilon)=e^{i\varphi}, \ \ \ \ x\in \mathbb{R}_{>4}
\label{a161}
\eeq
where $\nu_0$ is a parameter of the map. 
Under this map, the physical region above the cut $\nu>4$ is mapped to the upper half circle with angle $\varphi$:
\beq
x= \nu_0+\frac{8-2\nu_0}{1+\cos\frac{\varphi}{2}} 
\label{a159}
\eeq
Upon this map, we can then discretize the amplitude, form factors and spectral densities using the discrete basis of function \cite{Cordoba:2025wnv}.

The amplitude \eqref{amplitude} is parameterized by the constant $f_0$, and the single and double discontinuities $\sigma_{1,2},\rho_{1,2}$.  
The latter two can be expanded in a basis of functions $\phi_n (x) =\sin(n\varphi)$:\footnote{This parameterization was previously used in the S-matrix bootstrap context in \cite{He:2021eqn} and is analogous to the $\rho$-ansatz used in \cite{Paulos:2017fhb}.}
	\beq
	\sigma(x) = \sum_{n=1}^\infty \sigma_n \phi_n(x)  ,\; \rho(x,y) = \sum_{n,m=1}^\infty \rho_{nm} \phi_{n}(x) \phi_{m}(y) \label{a157}
	\eeq
These functions on the upper half circle, or equivalently on the real axis $x>4$ define analytic functions $\Phi_n(\nu)$ on the unit disk
\beq
\Phi_{n}(\nu) = \frac{1}{\pi} \int_4^\infty\!\!\!\!\!\! dx\ \frac{\phi_{n}(x)}{x-\nu}= z(\nu) ^n
\label{a160}
\eeq
and
\begin{subequations}
	\beqa
	\hat{\Phi}_{n,\ell}(s) &=& \int_{-1}^{+1} d\mu P_\ell(\mu) \Phi_{n}(t)  \label{a164}\\
	\tilde{\Phi}_{nm,\ell}(s) &=& \int_{-1}^{+1} d\mu  P_\ell(\mu) \Phi_{n}(t)\Phi_{m}(u)  \label{a165}
	\eeqa
\end{subequations}
where $P_\ell(\mu)$ are Legendre polynomials. The analytic amplitudes are written as
\beq\label{Astu}
\begin{aligned}
A(s,t,u) = &f_0 + \sum_{n=1}^\infty \left[\sigma^{(1)}_n \Phi_n(s) + \sigma^{(2)}_n (\Phi_n(t)+\Phi_n(u))\right]\\
+&\sum_{n,m=1}^\infty \left[\rho^{(1)}_{nm} \Phi_n(s) (\Phi_m(t)+\Phi_m(u)) + \rho^{(2)}_{nm} \Phi_n(t)\Phi_m(u) \right]
\end{aligned}
\eeq
with the corresponding partial waves given by
\beq
\begin{aligned}
h^I_\ell(s) = &\frac{\pi}{4} \sqrt{\frac{s-4}{s}} \Biggl\{f_0 A^I_\ell + \sum_{n=1}^\infty \left[ A^I_{\ell n}(s)\sigma^{(1)}_n + B^I_{\ell n}(s) \sigma^{(2)}_n \right] \\
+ &\sum_{n,m=1}^\infty  \left[ A^I_{\ell,nm}(s)\ \rho^{(1)}_{nm} + B^I_{\ell,nm}(s)\ \rho^{(2)}_{nm}\right] \biggr\} \ ,  
\end{aligned}\label{a176}
\eeq
where
\beq
\begin{aligned}
	A^0_\ell &= 2(N+2)\delta_{\ell 0},  & A^1_\ell &= 0, & A^2_\ell &= 4\delta_{\ell 0}, \\
	A^0_{\ell,n} &= 2N\delta_{\ell 0}\Phi_n+ 2\hat{\Phi}_{n\ell}, & A^1_{\ell,n} &= 2 \hat{\Phi}_{n\ell}, & A^2_{\ell,n} &= 2 \hat{\Phi}_{n\ell}, \\
	B^0_{\ell,n} &= 4\delta_{\ell 0} \Phi_n + 2(N+1) \hat{\Phi}_{n\ell},  & B^1_{\ell,n} &=  -2\hat{\Phi}_{n\ell},  & B^2_{\ell,n} &= 4 \delta_{\ell 0} \Phi_n + 2 \hat{\Phi}_{n\ell}, 
\end{aligned}
\eeq
and
\begin{subequations}
	\beqa
	A^0_{\ell,nm} &=& 2N\Phi_n\hat{\Phi}_{m\ell}+ 2\Phi_m\hat{\Phi}_{n\ell}+2\tilde{\Phi}_{nm,\ell}   \ ,\label{a186} \\
	A^1_{\ell,nm} &=& 2 \Phi_m\hat{\Phi}_{n\ell}+2\tilde{\Phi}_{nm,\ell}   \label{a187} \ ,\\ 
	A^2_{\ell,nm} &=& 2 \Phi_m\hat{\Phi}_{n\ell}+2\tilde{\Phi}_{nm,\ell}   \label{a188} \ ,\\
	B^0_{\ell,nm} &=&  \Phi_n\hat{\Phi}_{m\ell}+\Phi_m\hat{\Phi}_{n\ell}+ N\tilde{\Phi}_{nm,\ell} \ ,\label{a189} \\
	B^1_{\ell,nm} &=&  -\Phi_n\hat{\Phi}_{m\ell}-\Phi_m\hat{\Phi}_{n\ell} \label{a190} \ ,\\
	B^2_{\ell,nm} &=&  \Phi_n\hat{\Phi}_{m\ell}+\Phi_m\hat{\Phi}_{n\ell}  \label{a191} \ .
	\eeqa
\end{subequations}
To discretize the variables for the numerical computations, the sums in \eqref{Astu} and \eqref{a176} are truncated to
\begin{equation}
	\sum_{n=1}^\infty \rightarrow \sum_{n=1}^{n_{\text{max}}}\;,\;\; \ \ \ \sum_{n,m=1}^\infty \rightarrow \sum_{n,m=1}^{n_{\text{max}}}\;.
\end{equation}

The parameterization of the form factors is analogous. After the map \eqref{a161}, the form factor is an analytic function on the unit disk and can be expanded in series of $z$ and truncated:
\begin{equation}
	F^{I}_{\ell}(z)=\sum_{n=1}^\infty b^{I}_{\ell,n}z^n\to \sum_{n=1}^{n_b}b^I_{\ell,n}z^n
\end{equation}
For the currents $j_{S,V,T}$ with the unit normalization, the parameters $b^I_{\ell,n}$ are subject to the constraints (in units where $m_{\pi}=1$)
\begin{equation}
	F^I_{\ell}\big(z(0)\big)=1
\end{equation}
In the numerical computation, the real and imaginary parts of the form factors are evaluated, using truncated sums, on the upper half circle corresponding to the physical region above the cut:
\begin{equation}\label{RIF}
	\text{Re}F^{I}_{\ell}(\varphi)=\sum_{n=1}^{n_{b}} b^{I}_{\ell,n}\cos(n\varphi),\;\;\text{Im}F^{I}_{\ell}(\varphi)=\sum_{n=1}^{n_{b}} b^{I}_{\ell,n}\sin(n\varphi)
\end{equation}

\smallskip

In the numerical computation, we parameterized the spectral density $\hat{\rho}^I_\ell$ where the kinematic factor is stripped off:
\begin{equation}
    \rho^I_\ell=\kappa_\ell(s)\hat{\rho}^I_\ell(s)
\end{equation}
where $\kappa_\ell(s)$ are the kinematic factors taking the same expressions as \eqref{rholow}.

\medskip

The partial waves $h^I_{\ell}$, real and imaginary parts of the form factors $F^I_\ell$ and the spectral densities $\rho^I_\ell$ in the physical region are evaluated at a discrete set of points $x_i,i=1,\ldots,M$ for imposing the positive semidefinite constraints \eqref{Bmt}, partial waves unitarity \eqref{unit}, asymptotic form factors \eqref{chiIL} and sum rules \eqref{sr}. (See appendix C of \cite{Cordoba:2026wsy} for details) This discrete set of points are taken to be equally spaced on the upper half circle according to the map \eqref{a159}, namely
\begin{equation}\label{xi}
    x_i=\nu_0+\frac{8-2\nu_0}{1+\cos\frac{\varphi_i}{2}}
\end{equation}
where
\begin{equation}
    \varphi_i=\left(j-\frac{1}{2}\right)\Delta_{\phi},\;\;j=1,2,\ldots, M,\;\;\Delta_{\phi}=\frac{\pi}{M}
\end{equation}

\subsection{Varying $\nu_0$ with $m_\pi$}

To introduce the UV constraints of the QCD sum rules and asymptotic form factors, we choose a scale $s_0\simeq2\text{ GeV}$ beyond which we assume perturbative QCD applies. This means that the spectral density and form factor respectively go to their perturbative QCD values above $s_0$ and we are interested in solving the strongly coupled physics below $s_0$. Since we use the pion mass $m_\pi$ as a unit for energy, denoting the value of 2 GeV in pion mass unit as
\begin{equation}
 \hat{s}_0(m_\pi)=\frac{s_0}{m_\pi^2}    
\end{equation}
the value of $\hat{s}_0$ decreases with increasing pion mass $m_\pi$. From \eqref{xi}, for a fixed parameter $\nu_0$ of the map \eqref{a161}, the number of discretized numerical points below $s_0$
\begin{equation}
x_i<\hat{s}_0,\;\; i=1,\ldots,n_0
\end{equation}
decreases with increasing pion mass $m_\pi$.

To ensure we have the same numerical resolution below $s_0$ for the computations with different $m_\pi$, we therefore change the mapping parameter $\nu_0$ with $m_\pi$ and keep the number of points $n_0$ below $s_0$ fixed. The parameter $\nu_0$ for varying $m_\pi$ is
\begin{equation}
    \nu_0(m_\pi)=4-\frac{\hat{s}_0(m_\pi)-4}{\hat{s}_0(m^{\text{phys}}_\pi)-4}(4-\nu_0(m^{\text{phys}}_\pi))
\end{equation}
where $m^\text{phys}_\pi$ is the physical pion mass. This means that the numerical discretization changes for each $m_q$.

In the computations of this paper, we have
\begin{equation}
    \nu_0(m^{\text{phys}}_\pi)=-20,\;\; M=50,\;\; n_0=40
\end{equation}

\printbibliography

\end{document}